\documentclass[prb,twocolumn,superscriptaddress,showpacs]{revtex4}
\usepackage{times}
\expandafter\let\csname equation*\endcsname\relax
\expandafter\let\csname endequation*\endcsname\relax
\usepackage{amsmath}
\usepackage{amssymb}
\usepackage{amsfonts}
\usepackage{graphicx}
\usepackage{pst-all}

\begin{document}

\title{A Time-Reversal Invariant Topological Phase at the Surface of a 3D Topological Insulator}

\author{Parsa Bonderson}
\address{Station Q, Microsoft Research, Santa Barbara, California 93106-6105, USA}
\author{Chetan~Nayak}
\address{Station Q, Microsoft Research, Santa Barbara, California 93106-6105, USA}
\address{Department of Physics, University of California, Santa Barbara, California 93106, USA}
\author{Xiao-Liang Qi}
\address{Department of Physics, Stanford University, Stanford, California 94305, USA}
\address{Station Q, Microsoft Research, Santa Barbara, California 93106-6105, USA}

\begin{abstract}
A 3D fermionic topological insulator has a gapless Dirac surface state protected by time-reversal symmetry and charge conservation symmetry. The surface state can be gapped by introducing ferromagnetism to break time-reversal symmetry, introducing superconductivity to break charge conservation, or entering a topological phase. In this paper, we construct a minimal gapped topological phase that preserves both time-reversal and charge conservation symmetries and supports Ising-type non-Abelian anyons. This phase can be understood heuristically as emerging from a surface $s$-wave superconducting state via the condensation of eight-vortex composites. The topological phase inherits vortices supporting Majorana zero modes from the surface superconducting state. However, since it is time-reversal invariant, the surface topological phase is a distinct phase from the Ising topological phase, which can be viewed as a quantum-disordered spin-polarized $p_x + i p_y$ superconductor. We discuss the anyon model of this topological phase and the manner in which time-reversal symmetry is realized in it. We also study the interfaces between the topological state and other surface gapped phases.
\end{abstract}

\pacs{71.10.Pm, 05.30.Pr}
\maketitle









%

\section{Introduction}

There has been a great deal of recent theoretical and experimental
activity on topological insulators (TIs)~\cite{kanemelegraphene,kanemelez2,bhz2006, fukanemele2007, moorebalents2007, Roy2009, qhz2008, ludwigclass,Hasan10,Qi11}, which are
band insulators that cannot be adiabatically transformed into
the vacuum state so long as time-reversal and
charge-conservation symmetries are maintained.
A TI is an example of a more general class
called symmetry-protected topological (SPT) phases~\cite{Chen11a,Chen11b,Kitaev11,Lu12},
which cannot be adiabatically transformed
into the vacuum state so long as a symmetry group $G$ is preserved.
At the 2D surface of a system in a 3D SPT phase,
one of the following three possibilities must occur:
(1) there are gapless excitations; (2) the symmetry group $G$ is broken;
or (3) as noted by Senthil and Vishwanath~\cite{Senthil13b},
the surface develops topological order. In the latter case,
the system cannot be adiabatically continued into
the vacuum state even if the symmetry group $G$ is not preserved.

For instance, if the 3D system is in the shape of a solid torus, so that
the topologically-ordered surface is on the boundary of the solid torus,
then the system will have several degenerate ground states, unlike a trivial
insulator such as the vacuum.

Consider, for the sake of concreteness, a 3D non-interacting fermion TI.
In scenario 1, the surface state consists of an odd number of time-reversal
invariant (TRI) gapless Dirac fermions. In scenario 2, the surface
of a 3D TI can be gapped by breaking either time-reversal $\mathcal{T}$ symmetry or charge conservation $\mathcal{Q}$. When $\mathcal{T}$-symmetry is broken, the surface enters a
gapped integer quantum Hall state~\cite{qhz2008} with Hall
conductivity $\sigma_{xy}=\pm\frac{1}{2}\,\frac{e^2}{2h}$.
When charge conservation is broken by an $s$-wave pairing field,
the surface enters a superconducting state~\cite{Fu08} which has a
Majorana zero mode in each $hc/2e$ vortex.
Though the superconducting state supports vortices with non-trivial braiding statistics, it is not a topological
phase -- since it is \emph{gapless} -- but is, instead, a quasi-topological phase~\cite{Bonderson12} exhibiting broken symmetry.
A theory realizing scenario 3 -- a topologically-ordered $\mathcal{T}$- and $\mathcal{Q}$-symmetric
surface state -- has not been previously constructed for the fermionic TI,
despite its being the first 3D SPT phase discovered.
In this paper, we construct a topological phase that respects these symmetries.

The phase that we construct respects time-reversal invariance and charge conservation.
We believe that it is a minimal theory supporting Ising-type non-Abelian anyons and respecting both
of these symmetries.
It can be realized as the gapped surface of a 3D system according to scenario 3,
but cannot be realized in a strictly 2D system. We start from the Fu-Kane surface superconducting state~\cite{Fu08}, in which time-reversal symmetry is preserved and charge conservation is broken by the pairing order parameter. If one type of vortex proliferates due to quantum fluctuations, the system can be driven into an insulating state
with restored charge conservation. We consider the case in which the vortex that
condenses has flux $4hc/e$, or $8$ times the flux quantum $hc/2e$.
Based on general arguments, such as adiabatic charge pumping, we obtain the charge, topological spin, and time-reversal transformation properties of the minimal surface theory consistent with several further assumptions discussed in
Section~\ref{sec:vortexcond}.
The resulting state is an analogue of the $\nu=\frac{1}{2}$ Moore-Read Pfaffian state~\cite{Moore91,Greiter92,Nayak96c,Bonderson11},
but with its charge and neutral sectors having opposite chiralities with respect to each other.
We discuss the realization of time-reversal symmetry in this state and
analyze the boundary line between this state and symmetry-breaking
gapped surface states. Finally, we consider generalizations.

Our result will be explained using the formalism of anyon
models, so we begin with a brief introduction to topological charge,
fusion rules, quantum dimensions, and topological twist factors,
using, as examples, two models relevant to our construction.
In Appendix~\ref{sec:anyon_models}, we give a more detailed discussion, focussing
on associativity relations, braiding, and the action of modular transformations
on the torus, i.e. on $F$-symbols, $R$-symbols, and $S$-matrices.

\section{Review of some relevant anyon models}
\label{sec:anyon-models}

Before studying the surface topologically-ordered state,
it is helpful to review the topological properties of two relevant
anyon models that will be useful for constructing the surface theory.
We will use these models to fix the notation and terminology used
in the rest of the paper.

We first review the Ising anyon model. This model governs the topological order of
the gapped non-Abelian phase of Kitaev's honeycomb lattice model~\cite{Kitaev06a} and
of a quantum-disordered $p_{x} + i p_{y}$ superconductor.
The Ising anyon model has three quasiparticle types, labeled by the respective topological charges:
$I$ (vacuum), $\sigma$ (non-Abelian anyon), and $\psi$ (fermion). The fusion algebra or ``fusion rules'' specifying what can result when topological charges are combined or split are given by
\begin{equation}
\label{eq:Ising_fusion_rules}
\begin{array}{rrr}
I \times A = A \times I = A, & & \psi\times \psi =I, \\
\psi \times \sigma = \sigma \times \psi = \sigma, & & \sigma \times \sigma = I+ \psi,
\end{array}
\end{equation}
where $A \in \mathcal{C}_{\text{Ising}} = \left\{ I , \sigma, \psi \right\}$. It follows that the quantum dimensions are $d_I = d_\psi = 1$ (as is the case for all Abelian
anyons) and $d_\sigma = \sqrt{2}$. The quantum dimensions characterize how the dimension of the low-energy state space increases asymptotically as quasiparticles of the respective charge type are added to the system, e.g. ${\cal H}_n$ of $n$ $\sigma$-quasiparticles has $\text{dim}({{\cal H}_n}) = 2^{\frac{n}{2}-1} \sim d_{\sigma}^{n}$ for $n$ large.
The topological twist factors associated with the braiding statistics of the quasiparticle types are $\theta_I =1$, $\theta_{\sigma} = e^{i \pi /8}$, and $\theta_{\psi}=-1$.  The twist factors are the phases that result when the respective quasiparticles are rotated by $2\pi$, and can thus be thought of as related to the ``spin'' values $s = 0$, $1/16$, and $1/2$ (defined modulo 1) via $\theta_a = e^{i 2 \pi s_a}$.
If the system is on a curved surface, the topological twist factors
parameterize the coupling of the quasiparticles to the curvature.

We will also need the $\mathbb{Z}_{N}^{(w)}$ anyon model,
which has $N$ quasiparticle types, labeled by topological charges: $0,1,2,\ldots,N-1$.
The fusion rules for these topological charges are
$j \times k = [j+k]_{N}$, where we introduce the notation $[j]_{N} = j (\text{mod } N)$.
This clarifies the $\mathbb{Z}_{N}$ in the naming convention of this anyon model.
As these are all Abelian anyons, they all have quantum dimension $d_j = 1$.
The superscript $(w)$ in the naming convention signifies that the twist factors are $\theta_j = e^{i \frac{2 \pi w}{N} j^2}$.
For $N$ odd, $w$ can take any integer value (though these are not all distinct theories). For $N$ even, $w$ can take any integer or half integer value.

The $\mathbb{Z}_{8}^{(1/2)}$ anyon model will be particularly significant in this paper.
Such a theory, which corresponds to a
U$(1)_8$ Chern-Simons theory, can arise in the quantum
Hall regime if electrons pair into charge $-2e$ bosons that
condense into a $\nu=1/8$ bosonic Abelian quantum Hall state
(which has filling $\nu=1/2$ with respect to the electrons),
as discussed by Halperin~\cite{Halperin83}. We will denote such
a quantum Hall state by $H$. The topological charge label $j$ of a quasiparticle in $H$
is equal its electric charge in units of $e/4$ modulo $2e$.

In addition to the quasiparticle types, fusion rules, quantum dimenions, and topological twist factors
given above, anyon models are characterized
by associativity $F$-symbols and braiding $R$-symbols, which specify multi-particle
braid group representations. Finally, the $S$-matrix governs the modular
transformation properties of ground states on the torus, i.e. the invariance
of the theory under diffeomorphisms that are not continuously connected
to the identity. The definitions and properties of these quantities are
discussed in Appendix~\ref{sec:anyon_models}.

We can construct new models by combining Ising anyons and
Abelian anyon models, such as the $\mathbb{Z}_8^{(1/2)}$ model
given above. As we will see in the next section, the symmetry-respecting topological phase at the surface of a
3D fermionic TI is described by such a combined theory.
Such anyon models are completely determined by their
quasiparticle types, fusion rules, quantum dimensions, and topological twist factors~\cite{Bonderson07b,BondersonIP}.
In other words, these quantities uniquely specify the anyon model
(up to choice of gauge) and allow us to generate the $F$-symbols and $R$-symbols.
Consequently, we will primarily focus on deducing the quasiparticle types, fusion rules,
quantum dimensions, and topological twist factors in this paper.

\section{Vortex condensation and the surface topological state}\label{sec:vortexcond}

\subsection{Strategy and assumptions}

We now return to the surface of 3D topological insulators. We first outline the procedure that we will use to derive the surface theory. We start from the Fu-Kane superconductor and characterize the topological properties of the vortices in a manner analogous to the description of topologically ordered states. (Although the superconducting state
is a broken symmetry state, it is a quasi-topological state~\cite{Bonderson12} and,
therefore, has many of the properties
of a topological phase.) We conclude that an $8\pi$ vortex (i.e. flux $4hc/e$)
is the minimal neutral boson that can be condensed without breaking any symmetry.
Using very general considerations, we determine the topological spin
of the quasiparticles that remain when $8\pi$-vortex condensation destroys
superconductivity. This enables us to obtain the resulting
time-reversal- and charge-conservation-preserving
surface topological theory, which we will denote by $X$.
Finally, we subject our result to a number of consistency checks.

We first consider the Fu-Kane superconducting phase, and assume that
it is a 2D superconductor with 2D algebraic long-range order. Although the
state is not topologically-ordered, $\pi$ vortices have Majorana
zero modes and well-defined projective
non-Abelian statistics~\cite{Teo10,Freedman11,freedman2011b,Bonderson12}.
When two $\pi$ vortices are fused or braided, they behave like the non-Abelian $\sigma$-quasiparticle in the Ising topological phase, except that the twist factor is not well-defined
due to gapless phase fluctuations and the the long-range interactions that they mediate.
A double vortex with vorticity $2\pi$ has no Majorana zero mode and is, thus, Abelian. Another key difference between vortices in the Fu-Kane superconducting state
and Ising anyons is that the former carry a well-defined vorticity, which is conserved during the fusion process. Consequently, vortices with vorticity $\pi$ and $-\pi$ are distinct, although both carry similar Majorana zero modes. We will denote the non-Abelian vortex with vorticity $(2n+1)\pi$ by $\tilde{\sigma}_{2n+1}$, and denote the Abelian vortices with vorticity $2n\pi$ by $\tilde{I}_{2n}$ and $\tilde{\psi}_{2n}$. We have added tildes $\tilde{~}$ to the labels, to distinguish them from the ``real" topological charges of quasiparticles in the topologically-ordered state. Here, $\tilde{I}_{0}$ denotes the trivial/vacuum quasiparticle and $\tilde{\psi}_{2n}$ is obtained by creating a Bogoliubov quasiparticle $\tilde{\psi}_0$ in the state $\tilde{I}_{2n}$.
These vortices satisfy the fusion rules
\begin{eqnarray}
\label{eq:FK_fusion_rules}
\tilde{I}_{2m} \times \tilde{A}_{n} &=& \tilde{A}_{n} \times \tilde{I}_{2m} = \tilde{A}_{2m+n}, \nonumber\\
\tilde{\psi}_{2m}\times\tilde{\psi}_{2n}&=& \tilde{I}_{2m+2n} , \nonumber\\
\tilde{\psi}_{2m} \times \tilde{\sigma}_{2n+1} &=& \tilde{\sigma}_{2m+1}\times\tilde{\psi}_{2n} = \tilde{\sigma}_{2n+2m+1} , \nonumber\\
\tilde{\sigma}_{2m-1} \times \tilde{\sigma}_{2n+1} &=& \tilde{I}_{2m+2n}+ \tilde{\psi}_{2m+2n}
\end{eqnarray}
where $A \in \mathcal{C}_{\text{Ising}}$. Consequently, we assign quantum dimensions
$d_{\tilde{I}_{2n}} = d_{\tilde{\psi}_{2n}} = 1$ and $d_{\tilde{\sigma}_{2n+1}} = \sqrt{2}$.

Now we need to determine which vortex type can condense to drive a superconductor-insulator transition. In an ordinary superconductor, $\pi$ vortex condensation leads to
a band insulator (which would require charge-density-wave order unless
there is an even number of electrons per unit cell of the lattice) and
$2\pi$ vortex condensation leads to a Mott insulator. In the case of
a Fu-Kane surface superconductor, one needs to be careful due to the
presence of Majorana zero modes in $\pi$ vortices. We consider
the case in which the vortex that condenses is a boson, meaning that
it is Abelian and has trivial self-statistics~\footnote{More generally, one may consider topological transitions induced by condensing non-Abelian quasiparticles~\cite{Bais09}. In this case, the topological charge $a$ of the condensing quasiparticle must satisfy the conditions that $\theta_a =1$ and that there exist a fusion channel $c$ of two $a$ charges such that $(R^{aa}_c)^2 =1$. While the topological twists and overall phases of braiding factors are not well-defined in the Fu-Kane superconductor phase, one can also argue that for Ising-type non-Abelian anyons (meaning Ising up to Abelian anyon factors) where these are well-defined, one cannot simultaneously satisfy these conditions. In any case, we will not consider such condensations in this paper.}. If the vortex $\tilde{I}_{2p}$ with vorticity $2p\pi$ condenses, then vortices with vorticity $n\pi$ and $(n+2p)\pi$ are identified, and the vortices with vorticity $n\pi$ with $n=0,1,...,2p-1$ become topological
quasiparticles in the resulting topological state. In the following we will
denote vorticity $n\pi$ by $n$ for simplicity.

The arguments in the next two subsections depend on
the following natural assumptions.

\emph{There is a finite number of quasiparticle types in $X$.}
This is the usual assumption in a topological phase, without which
the theory cannot be tightly-constrained by self-consistency.
In the present context, since there are only $2p$ possible
distinguishable vorticities modulo $2p$ when
a $2p\pi$ vortex condenses, this assumption is equivalent to the
assumption that, at most, a finite number of new quasiparticle
types result from vortex condensation.

\emph{When a $2p$-vortex condenses, the quantum dimensions of the remaining
$n=0,1,...,2p-1$ vorticity vortices do not change, and Abelian and
non-Abelian vortices do not hybridize.} Physically, this means that the Majorana zero modes remain stable during the $2p\pi$ vortex condensation, since the $2p\pi$ vortex is invisible to Majorana zero modes (and also to Bogoliubov quasiparticles). Based on this assumption, we must obtain a theory with quasiparticles with quantum dimensions $1$ and $\sqrt{2}$.
Each quasiparticle in $X$ has a well-defined electrical charge, due to charge conservation symmetry.

\emph{There is at most one non-Abelian quasiparticle type with a given value
of electrical charge.}
This is even stronger than the previous assumption: we are assuming
that no new non-Abelian quasiparticle types result from vortex condensation.
This is the least-justified of our assumptions and, in fact, a recently-constructed
theory for the surface of a fermionic topological insulator
\cite{Metlitski-unpublished,Wang-unpublished} does not obey this assumption. Although we have no guarantee
that a theory that satisfies this assumption can be realized on the surface of a conventional fermionic TI (i.e. the
state that is continuously connected to a TI of non-interacting fermions), we will proceed and see if such a
theory exists and respects time-reversal and charge-conservation symmetries.

\subsection{Abelian quasiparticles}

\begin{figure}[t]
\includegraphics[width=1.60in]{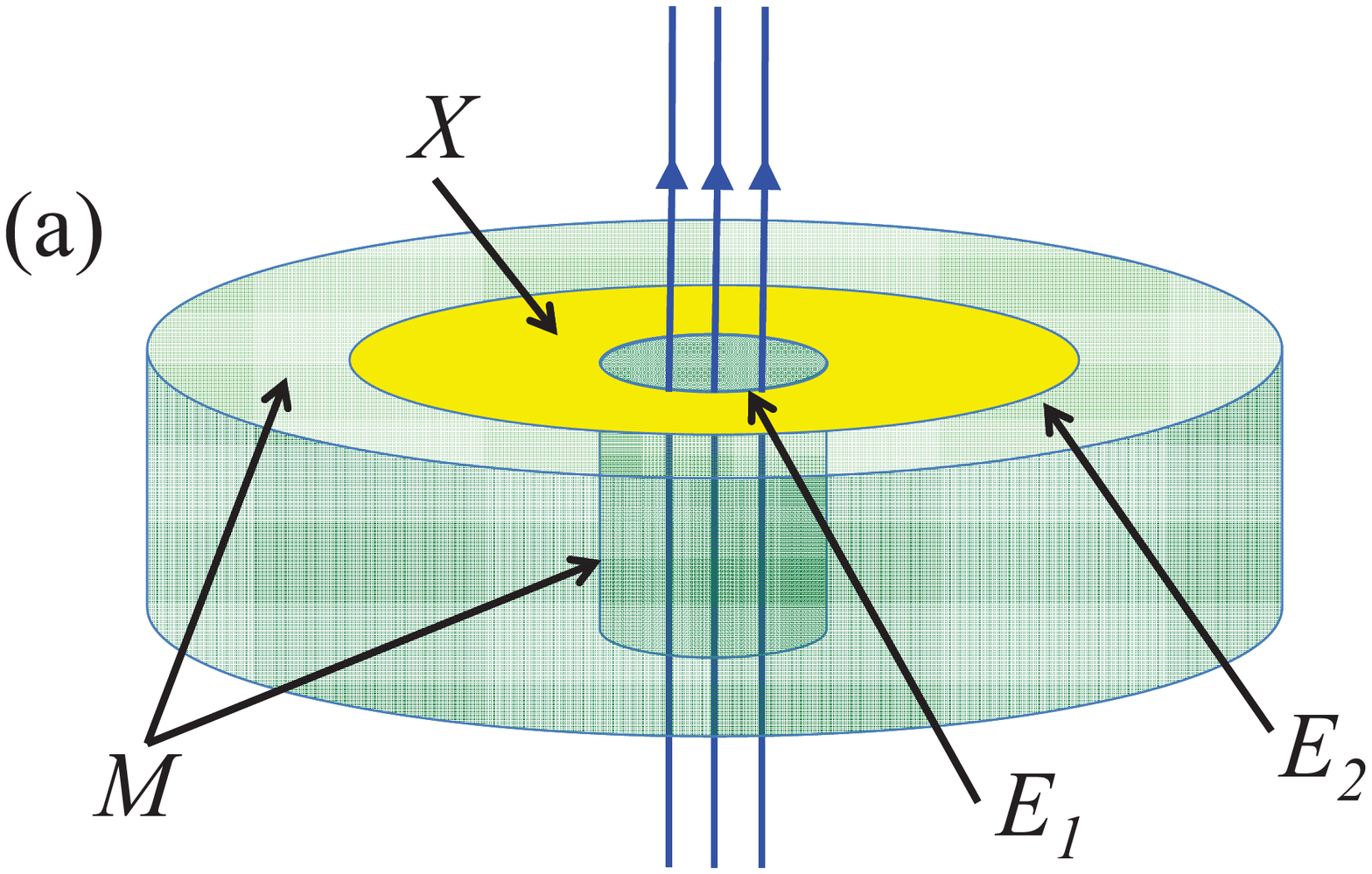}\hskip 0.1 in
\includegraphics[width=1.70in]{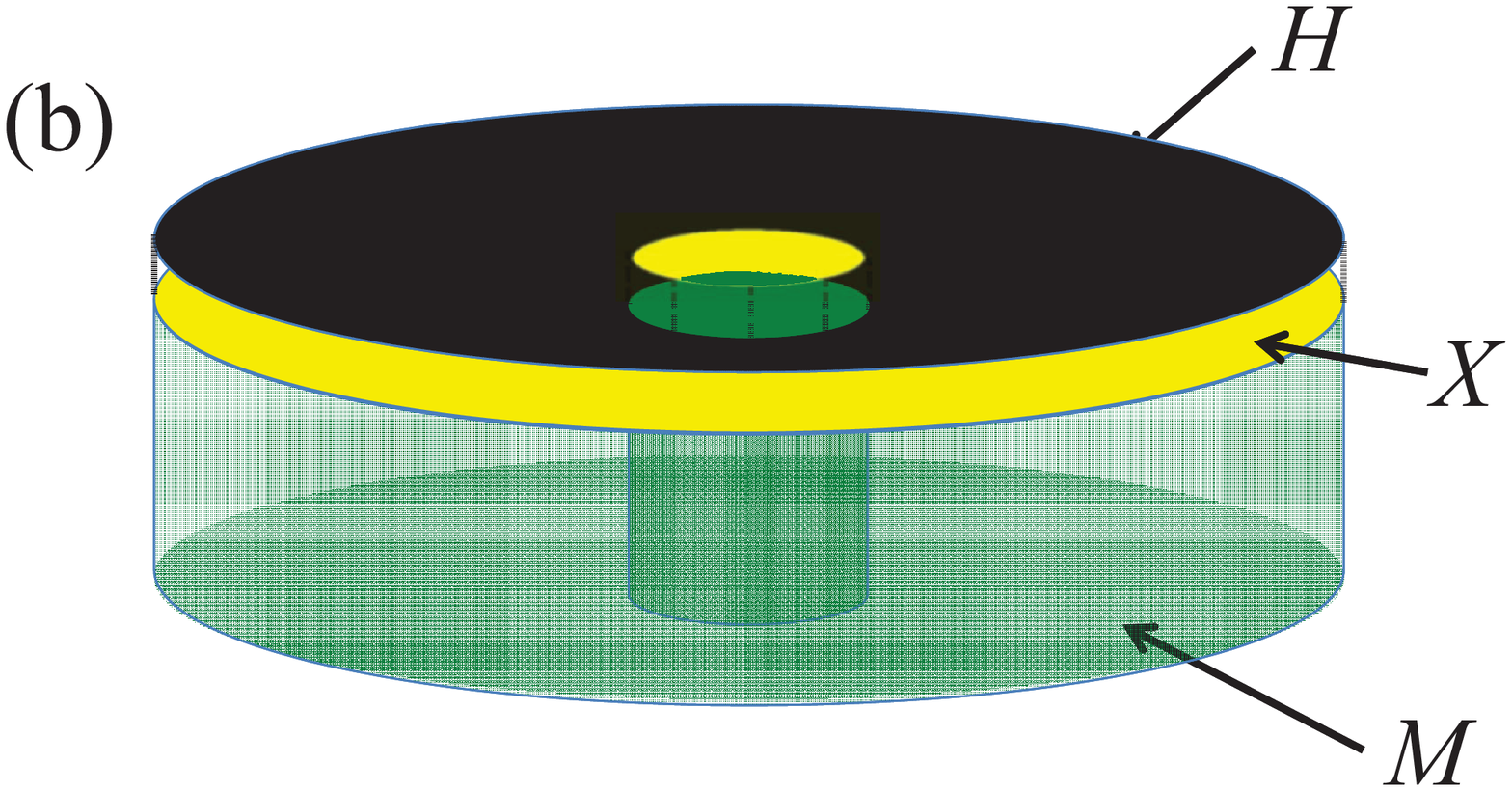}
\caption{(a) A toroidal slab of 3D TI has part of its upper surface in the gapped,
time-reversal-invariant and charge-conserving topological phase $X$
while the rest of it is in the IQH state $M$. When flux $hc/e$ is threaded
through the central hole, as shown, charge $e/2$ must accumulate
(b) The lower surface of a slab of 3D TI is in the phase $M$ and the upper surface
is in the phase $X$. In addition, we attach a 2D layer of electrons in the $\nu=1/2$
quantum Hall state $H$ of tightly-bound charge $-2e$ pairs in the $\nu_{\text {pair}}=1/8$ QH state
of the bosonic pairs. The combined system is a 2D system in the Ising topological phase,
and the excitation created by threading flux $hc/2e$ through the central tube is neutral.}
\label{fig:slab}
\end{figure}

We first focus on the Abelian quasiparticles which are descendants of
the $\tilde{I}_{2n}$ and $\tilde{\psi}_{2n}$ vortices in the quasi-long range
ordered phase. If a bosonic vortex $\tilde{I}_{2p}$ is condensed, we obtain Abelian quasiparticles with a well-defined (but possibly fractional) charge $q$ and a vorticity $2n$ modulo $2p$.
To understand the electrical charges carried by the Abelian quasiparticles, we consider an adiabatic flux-threading process in the configuration illustrated in Fig.~\ref{fig:slab} (a). The bulk of 3D TI is a solid torus, and part of the upper surface [the yellow annulus in Fig.~\ref{fig:slab}(a)] is in the symmetric topological phase $X$. The rest of the surface, denoted by $M$, is in a $\mathcal{T}$-breaking $\sigma_{xy}=\frac{e^2}{2h}$ integer quantum Hall (IQH) state with trivial topological order, which can be realized by coating the surface with a ferromagnetic insulator.
The states $X$ and $M$ occupy annular regions, which are joined together at
the edges $E_1$ and $E_2$, thereby forming a 2D torus on the surface of the 3D solid torus.

In this geometry, we can use Laughlin's adiabatic flux insertion argument~\cite{Laughlin83,Tao84}: by adiabatically
threading a magnetic flux through the central hole, we can transfer charge
(via the Hall conductance of $M$) from the inner edge $E_1$ to the outer edge
$E_2$, as shown in Fig.~\ref{fig:slab}(a). Since $X$ preserves time-reversal symmetry,
it does not contribute to the Hall current, and the charge pumped during this process is completely determined by the Hall conductance $\sigma_{xy} = e^2/2h$ of the $M$ region.
When $\Phi_0=hc/e$ flux is threaded through the central hole,
the charge pumped to the inner edge $E_1$ is $Q=\sigma_{xy} \Phi_{0}/c = e/2$. Since the system with flux $\Phi_0$ is gauge equivalent to the original system with flux $0$, the final state must be an energy eigenstate of the original Hamiltonian.
In other words, there must be charge $e/2$ quasiparticles in this system.
However, the phase $M$ is an IQH state with no fractionally-charged
excitations. Therefore, the state $X$ must have a charge $e/2$ quasiparticle excitation.
Meanwhile, threading flux $\Phi_0$ through the superconducting state
induces a vortex with vorticity $2\pi$ (corresponding to either $\tilde{I}_{2}$ or $\tilde{\psi}_2$).
Therefore it is natural to expect the charge $-e/2$ quasiparticle to
carry vorticity $2$ (modulo the yet-to-be determined period $2p$).

\begin{figure}
\centerline{
\includegraphics[width=3in]{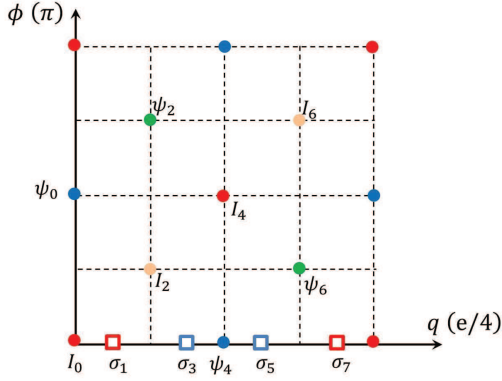}
}
\caption{\label{fig:qp} Quasiparticle types in theory $X$. The red hollow circles, blue solid circles, orange diamond and green diamond represent Abelian quasiparticles with
$\theta=1$, $i$, $-1$, and $-i$ respectively. The hollow squares stand for the non-Abelian quasiparticles. For Abelian quasiparticles, the $x$ and $y$ coordinates of the solid circles indicate the charge and vorticity, respectively, of the quasiparticle. The non-Abelian quasiparticles do not have well-defined vorticity but have a well-defined charge represented by their $x$ coordinates.
}
\end{figure}

We introduce the quantum numbers $(q,r)$ labeling quasiparticles with electric charge $q$ in units of $-e/4$ and
and vorticity $r$ in units of $\pi$. The quasiparticle types can be arranged on a grid, as shown in Fig.~\ref{fig:qp}.
In this notation, the quasiparticle obtained from the flux threading argument above is ascribed the label $(2,2)$.
The electron, which carries charge $-e$ and no vorticity, is ascribed the the label $(4,0)$.
By repeated fusion of these two Abelian quasiparticle types
$(2,2)$ and $(4,0)$, we can generate a lattice of Abelian quasiparticle types
\begin{eqnarray}
(2n+4m,2n) =(2,2)^n \times (4,0)^m
,
\end{eqnarray}
where fusion acts additively on the quantum numbers and the exponents indicate repeated fusion of the same quasiparticle type.
Since the fusion rules conserve electric charge and vorticity (before we consider the vortex condensation), each quasiparticle type is labeled by a well-defined electric charge and vorticity.
In principle, it is possible that the theory contains other Abelian quasiparticle types that are not generated by fusion of $(2,2)$ and $(4,0)$,
but the quasiparticle types $(2n+4m,2n)$ can be viewed
as the Abelian subtheory of a minimal possible surface theory.

Our next task is to determine the twist factors $\theta_{(2n+4m,2n)}$
for each quasiparticle type $(2n+4m,2n)$. The twist factors can be determined by the following three simple principles:
\begin{enumerate}
\item The trivial/vacuum quasiparticle corresponds to the label $(0,0)$. It follow that $\theta_{(0,0)}=1$ and all quasiparticles have trivial fusion and braiding with $(0,0)$. This is equivalent to the statement that a quasiparticle has the opposite electric charge and vorticity of its ``anti-quasiparticle,'' i.e. that $(-2n-4m,-2n)$ is the topological charge conjugate of $(2n+4m,2n)$. Since the twist factor of a topological charge and its conjugate are equal [see Eq.~(\ref{eq:twist})], this gives
\begin{equation}
\label{eq:cond_1}
\theta_{(2n+4m,2n)}=\theta_{(-2n-4m,-2n)}
.
\end{equation}
\item The electron $(4,0)$ is a fermion [i.e. $\theta_{(4,0)}=-1$] and is mutually local with all quasiparticles of $X$. In particular, this means the pure braid of $(2n+4m,2n)$ with $(4,0)$ is $R^{(2n+4m,2n)(4,0)}_{(2n+4m+4,2n)} R^{(4,0)(2n+4m,2n)}_{(2n+4m+4,2n)} = 1$. Using the ribbon property of Eq.~(\ref{eq:ribbon}), it follows that $\theta_{(2n+4m+4,2n)}=-\theta_{(2n+4m,2n)}$, and hence
\begin{equation}
\label{eq:cond_2}
\theta_{(2n+4m,2n)}=(-1)^{m} \theta_{(2n,2n)}
.
\end{equation}
\item The time-reversal transformation changes the sign of the vorticity, while preserving the electric charge, acting as $\mathcal{T} : (2n+4m,2n) \mapsto (2n+4m,-2n)$ on the quasiparticle labels, while complex conjugating the twist factors $\mathcal{T} : \theta_{(2n+4m,2n)} \mapsto \theta_{(2n+4m,-2n)}^{\ast}$. Therefore the time-reversal symmetry of the phase $X$ requires
    \begin{equation}
    \label{eq:cond_3}
    \theta_{(2n+4m,2n)}=\theta_{(2n+4m,-2n)}^{\ast}
    .
    \end{equation}
\end{enumerate}
Combining Eqs.~(\ref{eq:cond_1})-(\ref{eq:cond_3}), we obtain
\begin{equation}
\label{eq:cond_4}
\theta_{(2n,2n)}^{2} = (-1)^{n}
.
\end{equation}
In particular, $\theta_{(2,2)}=\pm i$. Since $\theta_{a^2} = \theta_{a}^4$ for any Abelian topological charge $a$ [see Eq.~(\ref{eq:Abelian_twists})], this gives us
\begin{equation}
\theta_{(2n+4m,2n)} = (-1)^{m} \left[ \theta_{(2,2)}\right]^{n^2}
\end{equation}
with $\theta_{(2,2)}=\pm i$. We emphasize that this has periodicity $2$ in both $m$ and $n$.

We note that the $(0,4)$ quasiparticle is an electrically neutral fermion [i.e. $\theta_{(0,4)} = -1$] and it is mutually local with all Abelian $(2n+4m,2n)$ quasiparticle types [since $R^{(2n+4m,2n)(0,4)}_{(2n+4m,2n+4)} R^{(0,4)(2n+4m,2n)}_{(2n+4m,2n+4)} = 1$]. Additionally, the lattice $(2n+4m,2n)$ of quantum numbers could equivalently be generated using $(2,2)$ and $(0,4)$, since $(4,0) = (2,2)^2 \times (0,4)^{-1}$.
We also notice that the $(0,8)$, $(8,0)$, and $(4,4)$ quasiparticles are all bosons [i.e. $\theta = 1$] and also mutually local with all $(2n+4m,2n)$ quasiparticle types. The $(0,8)$ bosonic quasiparticle is also electrically neutral.

If we assume that there is a finite number of quasiparticle types in $X$, then we must reduce the $(2n+4m,2n)$ lattice of quantum numbers by forming equivalence classes of quasiparticle types identified by fusion with some subset $\mathcal{Z} \subset \mathcal{C}$ of topological charges that act trivially in the theory (i.e. are bosonic and mutually local with all quasiparticles)~\cite{Bonderson07b}. In other words, we must specify a set $\mathcal{Z}$ of quasiparticle types that should be identified with the trivial/vacuum $(0,0)$ quasiparticle type to produce the reduced anyon model. Since the $(0,8)$, $(8,0)$, and $(4,4)$ quasiparticles generate all bosons in the $(2n+4m,2n)$ lattice, all the bosons are mutually local with all other $(2n+4m,2n)$ quasiparticles. Thus, we can specify $\mathcal{Z}$ by simply providing two linearly independent combinations of $(0,8)$, $(8,0)$, and $(4,4)$, as these will generate all elements of $\mathcal{Z}$. Consequently, the Abelian subtheory containing quasiparticle types $(2n+4m,2n)$ identified by $\mathcal{Z}$ must be described by an anyon model of the form
\begin{equation}
\mathcal{A} = \mathbb{Z}_{N_1}^{({N_1}/{2})} \times \mathbb{Z}_{N_2}^{(\pm {N_2}/{4})}
,
\end{equation}
where $N_1$ and $N_2$ are even, $\mathbb{Z}_{N_1}^{({N_1}/{2})}$ is generated by either $(4,0)$ or $(0,4)$, and $\mathbb{Z}_{N_2}^{(\pm {N_2}/{4})}$ is generated by $(2,2)$.

The minimal Abelian anyon model of this form is $\mathbb{Z}_{2}^{(1)} \times \mathbb{Z}_{2}^{(\pm 1/2)}$, which is obtained using $\mathcal{Z} = \left\{ (0,8) , (4,4) \right\}$ [or, equivalently, $\mathcal{Z} = \left\{ (8,0) , (4,4) \right\}$]. However, we will see (in the following subsection) that this anyon model is inadequate when one tries to include the non-Abelian quasiparticles.
Inclusion of non-Abelian anyons will require $N_2$ to be a multiple of $4$ and the minimal Abelian anyon model of this form is $\mathbb{Z}_{2}^{(1)} \times \mathbb{Z}_{4}^{(\pm 1)}$, which is obtained using $\mathcal{Z} = \left\{ (0,8) , (8,8) \right\}$ [or, equivalently, $\mathcal{Z} = \left\{ (0,8) , (8,0) \right\}$].
We can also arrive at this requirement by using the physical assumption that $\mathcal{Z}$ only contain the condensed bosonic vortex excitations and local bosonic excitations of the electronic system, i.e. quasiparticle types formed from an even number of electrons and/or holes, and quasiparticle types generated from these. This excludes $(4,4)$ from $\mathcal{Z}$, since it cannot be generated from bosonic excitations of this form (it requires the combination of an electron with a fermionic vortex excitation).

Since the $(0,8)$ quasiparticle, which can physically be viewed as a cluster of four $2\pi$ vortices $(2,2)$ together with two holes $(-4,0)$, is the electrically neutral boson with smallest vortcity, it is natural to use it for the vortex condensation.  This results in a non-trivial theory in which vorticity is well-defined modulo $8$. In this case, we should include $(0,8)$ in $\mathcal{Z}$ and it is most appropriate to write $\mathcal{A} = \mathbb{Z}_{2}^{(1)} \times \mathbb{Z}_{N_2}^{(\pm {N_2}/{4})}$, where the $\mathbb{Z}_{2}^{(1)}$ factor is the electrically neutral sector generated by the neutral fermion $(0,4)$. We emphasize that this results in exactly two Abelian quasiparticle types for a given electric charge value $q=2n$.

\subsection{Non-Abelian quasiparticles}

We now investigate the properties of the non-Abelian quasiparticles that descend from the non-Abelian $(2n+1)\pi$ vortices
$\tilde{\sigma}_{2n+1}$ in the superconducting state.
Before vortex condensation, we have the fusion rules
$\tilde{\sigma}_{2m-1}\times\tilde{\sigma}_{2n+1}=\tilde{I}_{2m+2n}+\tilde{\psi}_{2m+2n}$.
After vortex condensation, electric charge conservation must be preserved.
If we assume that no additional Abelian quasiparticle types are introduced and that $(0,8)$ vortex condensation occurred, so there are only two distinct Abelian quasiparticle types for a given value of electric charge $q=2n$, then the $\sigma$ quasiparticles must carry electric charge values $2j+1$ (in units of $-e/4$) with $j\in \mathbb{Z}$. It is not obvious how many distinct non-Abelian quasiparticles exist for a given value of electric charge, but we will make the minimal assumption that there is only one non-Abelian quasiparticle type for a given value $2j+1$ and denote it by $\sigma_{2j+1}$. With these assumptions, their fusion rules are completely determined to be
\begin{eqnarray}
\sigma_{2j+1}\times \sigma_{2k-1}&=& (2j+2k,2j+2k) \notag \\
&& \quad + (2j+2k,2j+2k+4)
\end{eqnarray}
and
\begin{eqnarray}
\sigma_{2j+1} \times (2n ,2n +4m) &=& (2n ,2n +4m) \times \sigma_{2j+1} \notag\\
&=& \sigma_{2j+2n+1}
.
\end{eqnarray}

{}From these fusion rules, we see that the fusion of two $\sigma$ quasiparticles can lead to Abelian quasiparticles with different vorticity, since there is only one quasiparticle type with given charge and vorticity. The two vorticity channels that may occur in the fusion of a pair of $\sigma$ quasiparticles always differ by a neutral fermion $(0,4)$ quasiparticle.
This is consistent with the view that the two fusion channels of $\pi$ vortices
differ by a neutral fermion excitation, just as the two states of a pair of Majorana zero modes
differ in fermion parity. Since the non-Abelian quasiparticles can fuse to different vorticity channels, they cannot be eigenstates of vorticity, even in the modulo $8$ sense.

We now determine the twist factors of $\sigma_{2j+1}$, similar to how we obtained the Abelian quasiparticles' twist factors.
\begin{enumerate}
\item The trivial/vacuum quasiparticle corresponds to the label $(0,0)$ requires $\sigma_{-2j-1}$ and $\sigma_{2j+1}$ to be topological charge conjugates of each other, and thus
    \begin{equation}
    \label{eq:cond_1_NA}
    \theta_{\sigma_{2j+1}}=\theta_{\sigma_{-2j-1}}
    .
    \end{equation}
\item The electron $(4,0)$ is a fermion and is mutually local with all quasiparticles of $X$. In particular, this means the pure braid of $\sigma_{2j+1}$ with $(4,0)$ is $R^{\sigma_{2j+1}(4,0)}_{\sigma_{2j+5}} R^{(4,0)\sigma_{2j+1}}_{\sigma_{2j+5}} = 1$. Using the ribbon property of Eq.~(\ref{eq:ribbon}), it follows that
    \begin{equation}
    \label{eq:cond_2_NA}
    \theta_{\sigma_{2j+5}}=-\theta_{\sigma_{2j+1}}
    .
    \end{equation}
\item The time-reversal transformation preserves the electric charge, which requires that $\mathcal{T} : \sigma_{2j+1} \mapsto \sigma_{2j+1}$ and $\mathcal{T} : \theta_{\sigma_{2j+1}} \mapsto \theta^{\ast}_{\sigma_{2j+1}}$.
    Therefore the time-reversal symmetry of the phase $X$ requires
    \begin{equation}
    \label{eq:cond_3_NA}
    \theta_{\sigma_{2j+1}}=\theta^{\ast}_{\sigma_{2j+1}}
    .
    \end{equation}
\end{enumerate}
Combining Eqs.~(\ref{eq:cond_1_NA})-(\ref{eq:cond_3_NA}), we obtain
\begin{eqnarray}
\theta_{\sigma_{8j+1}} &=& \theta_{\sigma_{8j+7}} = \theta_{\sigma_{1}} \\
\theta_{\sigma_{8j+3}} &=& \theta_{\sigma_{8j+5}} = -\theta_{\sigma_{1}}
,
\end{eqnarray}
where $\theta_{\sigma_{1}}=\pm 1$.

\begin{table*}[t!]
\begin{tabular}{ |c||c|c|c|c|c|c|c|c|c|c|c|c|}
  \hline
charge and vorticity labels&$(0,0)$ &$(0, 4) $& $(2,2)$&$(2,6)$&$(4,4)$ &$(4,0)$ &$(6,6)$&$(6,2)$&$\sigma_1$&$\sigma_3$&$\sigma_5$&$\sigma_7$\\
  \hline
${\rm Ising}\times \mathbb{Z}_8|_{\mathcal{C}}$ labels&$I_0$&$\psi_0$&$I_2$&$\psi_2$&$I_4$&$\psi_4$&$I_6$&$\psi_6$&$\sigma_1$&$\sigma_3$&$\sigma_5$&$\sigma_7$\\
  \hline
Time reversal &$I_0$&$\psi_0$&$\psi_2$&$I_2$&$I_4$&$\psi_4$&$\psi_6$&$I_6$&$\sigma_1$&$\sigma_3$&$\sigma_5$&$\sigma_7$\\
  \hline
Quantum dimensions &1&1&1&1&1&1&1&1&$\sqrt{2}$&$\sqrt{2}$&$\sqrt{2}$&$\sqrt{2}$\\
  \hline
   Twist factors for $w=-\frac{1}{2}$ & $1$ & $-1$ & $-i$ & $i$ & $1$ & $-1$ & $-i$ & $i$ & $1$ & $-1$ & $-1$ & $1$\\
  \hline
   Twist factors for $w=\frac{7}{2}$ & $1$ & $-1$ & $i$ & $-i$ & $1$ & $-1$ & $i$ & $-i$ & $-1$ & $1$ & $1$ & $-1$\\
   \hline
\end{tabular}
  \caption{Table of the $12$ quasiparticle types in the theory $X={\rm Ising}\times \mathbb{Z}_8^{(w)}|_{\mathcal{C}}$, where $w=-\frac{1}{2}$ and $\frac{7}{2}$, labeled by the electric charge and vorticity (first row) and by the $\left. \text{Ising}\times \mathbb{Z}_8 \right|_{\mathcal{C}}$ quasiparticle types (second row). The quasiparticle types that these transform into under time-reversal $\mathcal{T}$ are listed in the third row. The topological twist factors of the quasiparticle types for $w=-\frac{1}{2}$ and $\frac{7}{2}$ are listed in the fourth and fifth rows, respectively.}
 \label{table}
\end{table*}

\subsection{Anyon model $X$}

It is now convenient to change the notation of the Abelian quasiparticles. Since there are precisely two Abelian quasiparticle types at each value $2n$ of electric charge and they differ by a neutral fermion, we will denote the Abelian quasiparticles by $I_{2n} = (2n,2n)$ and $\psi_{2n} = (2n,2n+4)$.~\footnote{We note that there is some arbitrariness in how we assigned the labels of the semionic quasiparticles $I_{4n+2}$ and $\psi_{4n+2}$, and we are free to interchange these assignments, as long as we do it for all values of $n$, i.e. we could have instead defined the new labels as $I_{4n} = (4n,4n)$, $\psi_{4n} = (4n,4n+4)$, $I_{4n+2} = (4n+2,4n+6)$, and $\psi_{4n+2} = (4n+2,4n+2)$.} In this notation, the topological charges are $\mathcal{C} = \left\{ I_{2j}, \psi_{2j}, \sigma_{2j+1} | j =0,1,\ldots ,N-1 \right\} $, where $N$ is an even integer (still to be determined), and corresponding fusion rules are
\begin{eqnarray}
\label{eq:X_fusion_rules}
I_{2j} \times I_{2k} &=& I_{[2j+2k]_{2N}}, \notag \\
I_{2j} \times \psi_{2k} &=& \psi_{2k} \times I_{2j} = \psi_{[2j+2k]_{2N}}, \notag \\
I_{2j} \times \sigma_{2k+1} &=& \sigma_{2k+1} \times I_{2j} = \sigma_{[2j+2k+1]_{2N}}, \notag \\
\psi_{2j} \times \psi_{2k} &=& I_{[2j+2k]_{2N}}, \notag \\
\psi_{2j} \times \sigma_{2k+1} &=& \sigma_{2k+1} \times \psi_{2j} = \sigma_{[2j+2k+1]_{2N}}, \notag \\
\sigma_{2j+1} \times \sigma_{2k-1} &=& I_{[2j+2k]_{2N}}+ \psi_{[2j+2k]_{2N}},
\end{eqnarray}
which correspond to the restricted product of fusion algebras $ \left. \text{Ising} \times \mathbb{Z}_{2 N} \right|_{\mathcal{C}} $, using the shorthand notation $A_j \equiv (A,j)$ with $A \in \mathcal{C}_{\text{Ising}}$ and $j \in \{0,1,\ldots,2N-1\}$. Here, $N=N_2$ and the $\mathbb{Z}^{(1)}_{2}$ subsector generated by the neutral fermion $\psi_0 = (0,4)$ has been incorporated as the $\mathbb{Z}^{(1)}_{2}$ subsector of Ising.

The values of the topological twist factors $\theta_{\sigma_{2j+1}}$ constrain $N$ to be a multiple of $4$. In particular, $\sigma_{1}$ and $\sigma_{[-1]_{2N}}$ are required to be distinct from $\sigma_{[8j+3]_{2N}}$ and $\sigma_{[8j+5]_{2N}}$. If we take $N > 4$, we can see that the quasiparticle spectrum will simply repeat itself with every $2e$ electric charge, i.e. the quasiparticles will be indistinguishable under fusion with $I_8$, apart from the $2e$ difference in electric charge. Thus, we may as well identify the quasiparticle spectrum under fusion with $I_8$, i.e. taking $\mathcal{Z}=\left\{ (0,8),(8,8) \right\}$, and reduce to the minimal $N=4$ case.

Thus, we have constrained the minimal anyon model describe the phase $X$ to have the form
$ \left. \text{Ising} \times \mathbb{Z}^{(w)}_{8} \right|_{\mathcal{C}} $. It was found in Refs.~\onlinecite{Bonderson07b,BondersonIP} that anyon models of this form are uniquely distinguished by their topological twists~(see Appendix~\ref{sec:X_min_anyon_model}). In particular, the value of $\theta_{\sigma_{1}}$ determines the anyon model, with $\theta_{\sigma_{1}} = 1$ requiring $w = -\frac{1}{2}$ and $\theta_{\sigma_{1}} = -1$ requiring $w=\frac{7}{2}$ (the values of $w$ are modulo $8$). We notice that these choices of $w$ are consistent with the (already determined) Abelian subsector $\mathcal{A}$.

In summary, following the procedure above, we have determined the minimal anyon model representing the surface topological phase $X$ to be
\begin{equation}
X = \left. \text{Ising} \times \mathbb{Z}^{(w)}_{8} \right|_{\mathcal{C}}
\end{equation}
with $w = -\frac{1}{2}$ or $w=\frac{7}{2}$. These have $12$ quasiparticle types
\begin{equation}
\mathcal{C} = \left\{ I_{2j},\psi_{2j},\sigma_{2j+1} | j =0,1,2,3 \right\}
.
\end{equation}
The fusion rules are given
\begin{eqnarray}
\label{eq:Xmin_fusion_rules}
I_{2j} \times I_{2k} &=& I_{[2j+2k]_{8}}, \notag \\
I_{2j} \times \psi_{2k} &=& \psi_{2k} \times I_{2j} = \psi_{[2j+2k]_{8}}, \notag \\
I_{2j} \times \sigma_{2k+1} &=& \sigma_{2k+1} \times I_{2j} = \sigma_{[2j+2k+1]_{8}}, \notag \\
\psi_{2j} \times \psi_{2k} &=& I_{[2j+2k]_{8}}, \notag \\
\psi_{2j} \times \sigma_{2k+1} &=& \sigma_{2k+1} \times \psi_{2j} = \sigma_{[2j+2k+1]_{8}}, \notag \\
\sigma_{2j+1} \times \sigma_{2k-1} &=& I_{[2j+2k]_{8}}+ \psi_{[2j+2k]_{8}}.
\end{eqnarray}
The quantum dimensions and twist factors are listed in Table~\ref{table}. The surface theory $X$ is invariant under time-reversal and conserves electrical charge. The construction indicates that the theory $X$ can be viewed as a combination of Ising anyons with an Abelian theory whose quasiparticle types $0,1,2,...,7$ correspond to their respective values of electrical charge modulo $8$ (in units of $-e/4$).

With these quasiparticle types, fusion rules, quantum dimensions, and topological twist factors
in hand, we can obtain the corresponding $F$-symbols and $R$-symbols of the theory $X$ (e.g. by solving the pentagon and hexagon identities).
They are simply products of those of the Ising and $\mathbb{Z}_8^{(w)}$ anyon models, as explained in Appendix~\ref{sec:X_min_anyon_model}.

\section{Physical consequences and consistency checks}

\subsection{Time-reversal}

Since the state $X$ is time-reversal invariant, all physical properties computed
within this state must be invariant under an anti-unitary transformation that
permutes the quasiparticle types. One subtlety is that, even if the
$F$-symbols and $R$-symbols
are not invariant under a transformation, it may still be the case
that the system is invariant, because there is gauge freedom in defining
these quantities. For instance, if $R^{ab}_c=i$, then this will not be
invariant under complex conjugation. However, if
$a$ and $b$ are different quasiparticle types, then only $R^{ab}_c R^{ba}_c$ (corresponding to a pure braid)
is observable and gauge-invariant. A direct calculation~\cite{Bonderson07b,BondersonIP} shows that, up to gauge freedom,
an anyon model with fusion rules $\left. \text{Ising} \times \mathbb{Z}_{8} \right|_{\mathcal{C}}$
is uniquely determined by the fusion rules $N_{ab}^c$, quantum dimensions
$d_a$, and topological spins $\theta_a$. In other words, if these quantities are preserved under a transformation,
then it is the same theory, even if the $F$-symbols and $R$-symbols do not match exactly, because there must exist a gauge transformation that will make them match exactly.
Therefore, if the time-reversal transformation
maps quasiparticle types $\mathcal{T}: a \mapsto a'$,
then the condition for the theory to be time-reversal invariant is that $d_{a'}=d_{a}$, $\theta_{a'}^{\ast}=\theta_{a}$,
and $N_{ab}^c = N_{a'b'}^{c'}$. (See Appendix~\ref{sec:time_reversal_anyon} for the action of time-reversal on an anyon model.)
Leaving aside, for the moment, the microscopic construction in Section~\ref{sec:vortexcond}, one can ask the general question of what is the consistent action of time-reversal in theory $X$. From the topological twist factors
listed in Table~\ref{table}, we see that there are two
possible such time-reversal transformations for $X$:
\begin{eqnarray}
{\cal T}&:& {I_2} \leftrightarrow {\psi_2} \notag \\
&& {I_6}\leftrightarrow{\psi_6}
\end{eqnarray}
with all other quasiparticle types mapping to themselves; and
\begin{eqnarray}
{\cal T}'&:& {I_2} \leftrightarrow {\psi_6} \notag \\
&& {I_6}\leftrightarrow{\psi_2} \notag \\
&& \sigma_1 \leftrightarrow \sigma_7 \notag  \\
&& \sigma_3 \leftrightarrow \sigma_5
\end{eqnarray}
with all other quasiparticle types mapping to themselves.

The two time-reversal transformations are related by a relabeling of the quasiparticles
$A_{j} \rightarrow A_{8-j}$, where $A=I, \sigma, \psi$, which is a symmetry of the anyon model $X$. If we do not consider charge conservation symmetry, ${\cal T}$ and ${\cal T}'$ have the same physical effect. When we consider the assignment of electric charge to $A_j$, we are free to efine either $Q_{A_j}=-je/4$ or $Q_{A_j}=(j-8)e/4$. Once a choice of charge assignment is made, the two time-reversal transformations ${\cal T}$ and ${\cal T}'$ are no longer equivalent. Since time-reversal should preserve electric charge, ${\cal T}$ is the natural time-reversal transformation.

We can also examine the transformation properties of the system under ${\cal T}^2$. We consider time-reversal transformations whose application to states may involve a gauge transformation $U$ as well as quasiparticle type relabeling, because such a gauge transformation will later be needed to allow the $F$-symbols and $R$-symbols to remain invariant under $\mathcal{T}$. For this, we can focus on the transformations of the fusion/splitting vector associated with trivalent fusion/splitting vertices
\begin{eqnarray}
\mathcal{T} \left| a,b;c \right\rangle &=& u^{a' b'}_{c'} \left| a',b';c' \right\rangle \\
\mathcal{T}^2 \left| a,b;c \right\rangle &=& \left(u^{a' b'}_{c'} \right)^{\ast} u^{a b}_{c} \left| a,b;c \right\rangle
\end{eqnarray}
where $u^{a b}_{c}$ are the phase factors associated with the gauge transformation (see Appendix~\ref{sec:anyon_models} for more details).
Following Refs.~\onlinecite{Levin12,Fidkowski13}, we assign projective representations of ${\cal T}^2$ locally to each quasiparticle type $a$, which amounts to ascribing a local $\mathcal{T}^2$ value of $\eta_a$ to type $a$ quasiparticles~\footnote{Typically, one may assume $\eta_a = \pm 1$, but we can derive our results without making this assumption.}. Since $\mathcal{T}^2$ commutes with $\mathcal{T}$, we have the condition that $\eta_{a^{\prime}} = \eta_{a}^{\ast}$. Compatibility of the local $\mathcal{T}^2$ assignments with fusion requires that these satisfy
\begin{equation}
\mathcal{T}^2 \left| a,b;c \right\rangle = \eta_a \eta_b \eta_c^{-1} \left| a,b;c \right\rangle
,
\end{equation}
which gives the requirement that
\begin{equation}
\eta_a \eta_b \eta_c^{-1} = \left(u^{a' b'}_{c'} \right)^{\ast} u^{a b}_{c}
.
\end{equation}
This immediately produces the relations (among others)
\begin{eqnarray}
1 &=& \eta_{I_0} = \eta_{I_4}^{2} \eta_{I_0}^{-1} = \eta_{\psi_4}^{2} \eta_{I_0}^{-1} \notag \\
&=& \eta_{\sigma_1} \eta_{\sigma_7} \eta_{I_0}^{-1} =  \eta_{\sigma_1} \eta_{\sigma_7} \eta_{\psi_0}^{-1} =\eta_{\sigma_3} \eta_{\sigma_5} \eta_{I_0}^{-1}  \notag\\
&=& \eta_{\sigma_3} \eta_{\sigma_5} \eta_{\psi_0}^{-1} = \eta_{\sigma_1} \eta_{\sigma_3} \eta_{I_4}^{-1} =  \eta_{\sigma_1} \eta_{\sigma_3} \eta_{\psi_4}^{-1}
\end{eqnarray}
\begin{eqnarray}
\eta_{I_2} \eta_{\psi_2} \eta_{\psi_4}^{-1} &=& \left(u^{\psi_2 I_2}_{\psi_4} \right)^{\ast} u^{I_2 \psi_2}_{\psi_4} \\
\eta_{I_6} \eta_{\psi_6} \eta_{\psi_4}^{-1} &=& \left(u^{\psi_6 I_6}_{\psi_4} \right)^{\ast} u^{I_6 \psi_6}_{\psi_4}
\end{eqnarray}
\begin{eqnarray}
\eta_{\sigma_1}^{2} \eta_{I_2}^{-1} &=& \eta_{\sigma_1}^{-2} \eta_{\psi_2} = \left(u^{\sigma_1 \sigma_1 }_{\psi_2} \right)^{\ast} u^{\sigma_1 \sigma_1}_{I_2} \\
\eta_{\sigma_3}^{2} \eta_{I_6}^{-1} &=& \eta_{\sigma_3}^{-2} \eta_{\psi_6} = \left(u^{\sigma_3 \sigma_3 }_{\psi_6} \right)^{\ast} u^{\sigma_3 \sigma_3}_{I_6}
,
\end{eqnarray}
which reduce to
\begin{eqnarray}
\eta_{I_0} &=& \eta_{\psi_0} = 1 \\
\eta_{\sigma_1} &=& \eta_{I_4} \eta_{\sigma_3} = \eta_{I_4} \eta_{\sigma_5} = \eta_{\sigma_7} \\
\eta_{I_4} &=& \eta_{\psi_4} =  \left(u^{I_2 \psi_2}_{\psi_4} \right)^{\ast} u^{\psi_2 I_2}_{\psi_4}  \\
\eta_{I_2} &=& \eta_{\psi_2}^{\ast} = \left(u^{\sigma_1 \sigma_1 }_{I_2} \right)^{\ast} u^{\sigma_1 \sigma_1}_{\psi_2} \label{eq:eta_I_2} \\
\eta_{I_6} &=& \eta_{\psi_6}^{\ast} = \left(u^{\sigma_3 \sigma_3 }_{I_6} \right)^{\ast} u^{\sigma_3 \sigma_3}_{\psi_6} \label{eq:eta_I_6} \\
\eta_{I_4}^2 &=& \eta_{\sigma_1}^2 = \left| \eta_{I_2} \right| = \left| \eta_{I_6} \right| = 1
.
\end{eqnarray}

We can now impose the condition that $\mathcal{T}$ commutes with fusion and braiding, in the sense that the $F$-symbols and $R$-symbols are invariant under time-reversal. This imposes some constraints on the gauge transformation $U$ incorporated in $\mathcal{T}$, but it is assured that this is always possible, since the anyon model $X$ is invariant (up to gauge transformation) under time-reversal. The main constraint that we will use comes from braiding $I_2$ and $\psi_2$. Eq.~(\ref{eq:ribbon}) indicates that
\begin{equation}
R^{I_2 \psi_2}_{\psi_4} R^{\psi_2 I_2}_{\psi_4} = \frac{\theta_{\psi_4} }{ \theta_{I_2} \theta_{\psi_2}} =-1
,
\end{equation}
and thus
\begin{equation}
R^{I_2 \psi_2}_{\psi_4} = - \left[ R^{\psi_2 I_2}_{\psi_4} \right]^{\ast}
.
\end{equation}
Note that this relation is gauge invariant. Under time-reversal, the $R$-symbols transform as
\begin{equation}
\mathcal{T} : R^{I_2 \psi_2}_{\psi_4} \mapsto \frac{u^{\psi_2 I_2}_{\psi_4} } { u^{I_2 \psi_2 }_{\psi_4} } \left[ R^{\psi_2 I_2}_{\psi_4} \right]^{\ast}
,
\end{equation}
so for this to leave $R^{I_2 \psi_2}_{\psi_4}$ unchanged, the associated gauge transformation is required to have
\begin{equation}
\left( u^{I_2 \psi_2 }_{\psi_4} \right)^{\ast} u^{\psi_2 I_2}_{\psi_4} = -1.
\end{equation}
Using this, we obtain
\begin{eqnarray}
\eta_{I_0} &=& \eta_{\psi_0} = +1 \\
\eta_{I_4} &=& \eta_{\psi_4} = -1  \\
\eta_{\sigma_1} &=& - \eta_{\sigma_3} = - \eta_{\sigma_5} = \eta_{\sigma_7} = \pm 1 .
\end{eqnarray}
We emphasize that these results are independent of the choice of gauge transformation $U$ incorporated in $\mathcal{T}$, beyond the fact that it is chosen to leave the $R$-symbols invariant under $\mathcal{T}$. We also note that the sign of $\eta_{\sigma_1}$ is not constrained by any choice of $U$, because $\sigma_j$ always enter fusion vertices in pairs. The values of $\eta_{I_2}$ and $\eta_{I_6}$ appear to be phases that depend on the choice of gauge transformation $U$, as indicated in Eqs.~(\ref{eq:eta_I_2}) and (\ref{eq:eta_I_6}) (though it might be the case that these are constrained by the requirement that $\mathcal{T}$ leave the $F$-symbols unchanged).

Thus, we have found that, for a time-reversal transformation $\mathcal{T}$ that leaves the $F$-symbols and $R$-symbols invariant, the vacuum $I_0$ and neutral fermion $\psi_0$ transform as $\mathcal{T}^2 =+1$, while the electron $\psi_4$ and charge $e$ boson $I_4$ transform as $\mathcal{T}^2 =-1$.

\subsection{Topology and modular transformations}

A defining characteristic of topological phases is how they behave when the system has nontrivial topology and when it is acted upon by topological operations. This behavior is encoded in a corresponding topological quantum field theory (TQFT) and can be reduced to the representation of modular transformations.
The anyon models $X = \left. \text{Ising} \times \mathbb{Z}^{(w)}_{8} \right|_{\mathcal{C}}$ with $w = -\frac{1}{2}$ or $w=\frac{7}{2}$ are not modular in the usual sense, because the $S$-matrix is degenerate. We can, however, apply a $\mathbb{Z}_2$-grading to this theory to produce a theory that is ``spin modular'' (see Appendix~\ref{sec:modularity_anyon}; this is identical to the case of the $\nu=1/2$ Moore-Read Pfaffian FQH state). For this, we form $\mathbb{Z}_2$ doublets of quasiparticle types under fusion with the electron quasiparticle type, $\widehat{a} = \{ a , a \times \psi_4 \}$, giving the unitary $\mathbb{Z}_2$-graded $\widehat{S}$-matrix
\begin{equation}
\widehat{S} = \frac{1}{\sqrt{8}} \left[
\begin{array}{cccccc}
1  &  1  &  \sqrt{2} &  1  &  1  &  \sqrt{2} \\
1  &  1  &  -\sqrt{2} &  1  &  1  &  -\sqrt{2} \\
\sqrt{2}  &  -\sqrt{2}  & 0  &  -i \sqrt{2}  &  i \sqrt{2}  & 0 \\
1  &  1  &  -i \sqrt{2} &  -1  &  -1  &  i \sqrt{2} \\
1  &  1  &  i \sqrt{2} &  -1  &  -1  &  -i \sqrt{2} \\
 \sqrt{2}  &   - \sqrt{2}&  0 &  i \sqrt{2}   &  -i \sqrt{2}   &  0
\end{array}
\right]
\end{equation}
where the order of the rows and columns is $\widehat{I}_0$, $\widehat{\psi}_0$, $\widehat{\sigma}_1$, $\widehat{I}_2$, $\widehat{\psi}_2$, and $\widehat{\sigma}_3$.

In this way, the anyon model has a corresponding $\mathbb{Z}_2$-graded TQFT (or ``spin field theory'') that allows one to describe topological properties and operations when the system occurs on manifolds that permit spin structures. For such a theory, one must specify the manifold and a choice of spin structure, to which the $\mathbb{Z}_2$-graded TQFT associates a Hilbert space. In order for modular transformations to map states back into the same Hilbert space, one must restrict to modular transformations that preserve the spin structure.

For example, if we consider a topological phase described by the anyon models $X$ on a 2D torus with no quasiparticle excitations,
then, with a permissible choice of spin structure (which is essentially the choice of periodic or antiperiodic boundary conditions around the two cycles of the torus), the ground state Hilbert space will be $6$-dimensional, corresponding to the number of $\mathbb{Z}_2$-graded quasiparticle type doublets. The $\widehat{S}$-matrix represents the modular transformation that interchanges cycles of a torus (when there are no quasiparticles).

The system can also be defined on a torus when there is a single $\psi_0$ quasiparticle (or all the quasiparticles are in the collective fusion channel $\psi_0$). In this case, the state space is $2$-dimensional and interchanging cycles of a torus is represented by
\begin{equation}
\widehat{S}^{(\psi_0)} = \frac{-i}{\sqrt{2}} \left[
\begin{array}{cc}
1  &  -i  \\
-i  &  1
\end{array}
\right]
\end{equation}
where the order of the rows and columns is $\widehat{\sigma}_1$ and $\widehat{\sigma}_3$.

Another significant modular transformation is the Dehn twist, which is obtained by cutting open the torus along a cycle and gluing the resulting boundaries back together with a $2\pi$ twist. For a TQFT, this operations is represented by the $T$-matrix. For a $\mathbb{Z}_2$-graded TQFT, $\widehat{T}$ is not unambiguously defined, because the twist factors of the two quasiparticle types in a $\mathbb{Z}_2$ doublet have opposite sign, however, the double Dehn twist (cutting open the torus along a cycle and gluing the resulting boundaries back together with a $4\pi$ twist) can be unambiguously defined by $\widehat{T}^2$. For the $\mathbb{Z}_2$-graded TQFT corresponding to anyon model $X$,
\begin{equation}
\widehat{T}^2 = \text{diag} \left[ 1, 1, 1, -1, -1, 1 \right]
\end{equation}
represents the action of the double Dehn twist.

We can also ask whether the topological phase $X$ could arise as a strictly 2D system. For this, we consider gauging the fermion parity symmetry or, in other words, obtaining the $\mathbb{Z}_2$-graded modular extensions of the anyon model $X$~\cite{Gu13,Fidkowski13}. The $\mathbb{Z}_2$-graded modular extensions of $X = \left. \text{Ising} \times \mathbb{Z}^{(w)}_{8} \right|_{\mathcal{C}}$ are described in Appendix~\ref{sec:modular_extensions}. In order to respect time-reversal symmetry, we only consider those with central charge $c_{-}=0$. These have the form~\footnote{We thank M. Metlitski for pointing these out to us.}
\begin{equation}
\tilde{X} = \frac{\text{Ising}^{(m+n)} \times \text{Ising}^{(n)} \times \mathbb{Z}_8^{(w-n)}}{\left\langle (f,\psi,4) \right\rangle}
,
\end{equation}
where $(m,n)=(0,0)$, $(4,1)$, $(4,2)$, or $(0,3)$ for $w=-1/2$ and $7/2$, and $\text{Ising}^{(n)}$ denotes the $n$th Galois conjugate of the Ising anyon model (see Appendix~\ref{sec:Ising_anyons}). The denominator indicates moding out by or condensing the boson $(f,\psi,4)$ in the direct product $\text{Ising}^{(m+n)} \times \text{Ising}^{(n)} \times \mathbb{Z}_8^{(w-n)}$.
The resulting anyon model has 18 distinct topological charges
\begin{equation}
\tilde{\mathcal{C}} = \left\{
I_{2j} , \psi_{2j} , \sigma_{2j+1} , s_{2j+1} , s\sigma_{0}, s\sigma_{2}
| j=0,1,2,3  \right\}
.
\end{equation}
The fusion and braiding properties are simply given by the product of those of $\text{Ising}^{(m+n)} \times \text{Ising}^{(n)} \times \mathbb{Z}_8^{(w-n)}$, up to the identification by fusion with the boson. For example, the extending quasiparticle types $s_{2j+1}$, $s \sigma_0$, and $s \sigma_2$ have fusion rules
\begin{eqnarray}
s_{2j+1} \times s_{2k-1} &=& I_{[2j+2k]_{8}}+ \psi_{[2j+2k+4]_{8}} \\
s\sigma_0 \times s\sigma_0 &=& s\sigma_2 \times s\sigma_2 = I_0 + \psi_0 +I_4 +\psi_4
\end{eqnarray}
and quantum dimensions $d_{s_{2j+1}} = \sqrt{2}$ and $d_{s \sigma_{0}}=d_{s \sigma_{2}}=2$. The corresponding topological twist factors have the four possible combinations of
\begin{eqnarray}
\theta_{s_1} &=& -\theta_{s_3} = -\theta_{s_5}=\theta_{s_7}=\pm 1 \\
\theta_{s\sigma_0} &=& \theta_{s\sigma_2}^{\ast} = e^{\pm i \pi/4}
.
\end{eqnarray}

It should be clear that these modular extensions do not preserve electric charge conservation. They are time-reversal invariant, but not in a way that is compatible with the time reversal operation $\mathcal{T}$. Rather, they are compatible with the time reversal operation $\mathcal{T}^{\prime}$ acting as
\begin{eqnarray}
{\cal T}'&:& {I_2} \leftrightarrow {\psi_6} \notag \\
&& {I_6}\leftrightarrow{\psi_2} \notag \\
&& \sigma_1 \leftrightarrow \sigma_7 \notag  \\
&& \sigma_3 \leftrightarrow \sigma_5 \notag  \\
&& s_1 \leftrightarrow s_7 \notag  \\
&& s_3 \leftrightarrow s_5 \notag  \\
&& s\sigma_0 \leftrightarrow s\sigma_2
\end{eqnarray}
and all other quasiparticle types mapping to themselves.

Thus, we see that the topological phase $X$ cannot occur in a strictly 2D time-reversal and charge-conserving system, but, rather, is required to be the surface termination of a 3D time-reversal and charge-conserving system.

\subsection{Relation to Ising anyons}

Since we have not given an explicit derivation of our anyon model
from a microscopic theory of the surface of a fermionic TI,
it is useful to have a heuristic picture of this state.
To formulate such a picture, consider a 3D TI in a
toroidal geometry formed by taking an annular slab
on which the top surface is in the state $X$ and the bottom surface is
in the state $M$. To this system, we append a 2D layer of electrons
in the $\nu=1/2$ state, in which tightly-bound charge $-2e$ pairs
form a bosonic quantum Hall state at $\nu_{\text {pair}}=1/8$
(a state first suggested in Ref.~\onlinecite{Halperin83}).
This is denoted by $H$ in Fig.~\ref{fig:slab}(b). We have discussed in Section~\ref{sec:anyon-models} that $H$ corresponds to the anyon model $\mathbb{Z}_8^{(1/2)}$.
We can view this entire complex as a 2D system. (Suppose that the annular
slab has very large or infinite extent in the $x, y$ directions,
so that its $z$-thickness is much smaller than its extent in these directions.
In this limit, the entire system is 2D when probed on length scales much larger
than the thickness in the $z$-direction.)
Since the additional 2D layer cancels the Hall conductance of the lower surface
of the slab, this 2D system is in a topological phase with broken time-reversal invariance, vanishing (electrical) Hall conductance, and Majorana zero
modes. In our state $X$,  the entire complex in Fig.~\ref{fig:slab}(b)
is topologically equivalent to an Ising theory if we restrict the system
to electrically-neutral quasiparticles.

If we now insert flux through the central hole, we create quasiparticles
in both $X$ and $H$. Let $A_j$, $j$, and $A$, etc. describe
quasiparticles of the $X$, $H$, and Ising systems, respectively.
Then, quasiparticles $A$ of the Ising theory must be electrically-neutral
composites of $A_j$ in $X$ and $j$ in $H$. From this, it is natural to have
$X = \left. \text{Ising}\times \mathbb{Z}_8^{(-1/2)} \right|_{\mathcal{C}}$, as expected.

\subsection{Edge excitations}

A 2D state that is closely related to $X$ can be
obtained by taking the slab geometry in Fig.~\ref{fig:slab}(a)
and viewing the entire slab as a 2D system.\footnote{At length
scales much larger than the thickness of the slab in the $z$-direction,
we may treat the system as two-dimensional.}
(There are two such states, since we can choose either chirality
in the $M$ region, i.e. a gap of either sign.) This 2D state
has an edge to the vacuum, which is the outer cylinder of the
slab in Fig.~\ref{fig:slab}(a). (At length scales much larger than
the thickness of the slab, the cylinder can be treated as a circle.)
It has $\sigma_{xy}=-\frac{e^2}{2h}$ due to the bottom $M$ surface.
Therefore, the 2D system necessarily breaks time-reversal
and the time-reversal-invariant state $X$ necessarily lives only
on the surface of a 3D system. But since the bottom $M$ surface
is an integer quantum Hall state, it has no quasiparticles with non-trivial
braiding properties. Therefore, the state $X$ and its related
$\sigma_{xy}=-\frac{e^2}{2h}$ 2D topological state have the same
bulk quasiparticle fusion and braiding properties. The only difference
is in the realization of time-reversal. This is manifested in the edge
between the 2D state and the vacuum which can be deduced from
its Hall conductance and the relation of this state to the Ising anyon model
discussed in the previous subsection. The 2D state has
a chiral bosonic edge mode and a neutral Majorana fermion edge mode, as does the
Moore-Read Pfaffian state, but the two modes here are oppositely directed,
unlike in the Moore-Read state. The 2D state
has $\sigma_{xy}=-\frac{e^2}{2h}$
and chiral central charge
$c-\overline{c}=-\frac{1}{2}$,
which leads to a thermal Hall conductivity
$\kappa_{xy}=\frac{1}{2}\frac{{\pi^2}{k_B^2}T}{3h}$ at finite temperature.

\begin{figure}
\centerline{\includegraphics[width=3.5 in]{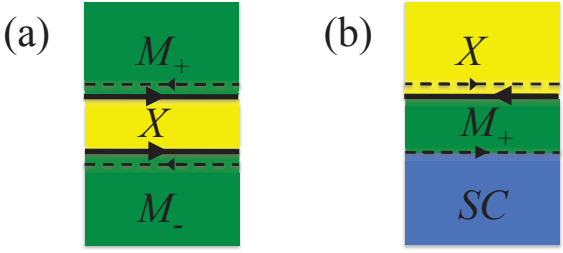}}
\caption{(a) A thin strip of $X$ separates $M_+$ and $M_-$,
in which time-reversal symmetry is broken oppositely. As this strip
becomes narrower, the two edges interact and form a single integer
quantum Hall edge.
(b) A thin strip of $M_+$ separates $X$ from $SC$.
As this strip becomes narrower, the two edges interact and form a critical
transverse field Ising model.}
\label{fig:edges}
\end{figure}

While $X$ cannot have an edge separating it from the vacuum,
it is possible for it to have an edge separating it from the state $M$
or the Fu-Kane superconductor, which we will denote $SC$.
The former edge has precisely the same
edge structure with $c-\overline{c}=-\frac{1}{2}$ as the edge
between the related 2D state discussed above and the vacuum,
as is clear from Fig.~\ref{fig:slab}(a). This edge can be understood
as one half of an integer quantum Hall edge as follows.
Consider the situation depicted in Fig.~\ref{fig:edges}(a):
A strip of phase $X$ separates regions of $M_+$ and $M_-$,
in which time-reversal symmetry is broken oppositely.
The effective theory of the two edges is:
\begin{multline}
S = \int \! dt \, dx \Bigl[\frac{2}{4\pi} \partial_x {\phi_1}
({\partial_t}+{v^c_1}\partial_x)\phi_1 +
i \psi_1({\partial_t}-{v_1}\partial_x)\psi_1\\
+\frac{2}{4\pi} \partial_x {\phi_2}
({\partial_t}+{v^c_2}\partial_x)\phi_2 +
 i \psi_2({\partial_t}-{v_2}\partial_x)\psi_2\\
+ {t_1} \Psi_R \Psi_L + \text{h.c.}
 + {t_2} \Psi^\dagger_R \Psi_L + \text{h.c.}\Bigr]
\end{multline}
Here, the subscripts $1$ and $2$ refer to the $M_{+}$-$X$
and $M_{-}$-$X$ edges [the upper and lower edges in Fig.~\ref{fig:edges}(a)]. Grassman fields $\psi_{1}$ and $\psi_{2}$ describe the left-moving neutral Majorana fermions, and the chiral boson fields $\phi_{1}$ and $\phi_{2}$ describe the right-moving charged bosons. Using the chiral bosons, we can form the charged fermion (electron) operator $f_R=e^{i(\phi_1+\phi_2)}$ and the neutral fermion operator $\Psi_R=e^{i(\phi_1 - \phi_2)}$. The couplings $t_1$ and $t_2$ couples the right-moving neutral fermion $\Psi_R$ to the left-moving fermion $\Psi_L=\psi_1+i\psi_2$ formed by the two Majorana fermions. As the $X$ strip is made narrower, the couplings $t_1$ and $t_2$ become
larger. 
These terms fully gap the neutral sector of the theory, which has $c=1$ from the neutral mode $\Psi_R$
and $\overline{c}=\frac{1}{2}+\frac{1}{2}$ from $\psi_1$, $\psi_2$.
Thus, all that is left is the $c=1$ total charge mode $f_R$, carrying
$\sigma_{xy}=1$ and $\kappa_{xy}=1$.

On the other hand, the edge separating $X$ from $SC$
need not, in principle, have any gapless excitations.
Since the states on both sides of the
edge preserve time-reversal symmetry, the edge will not have
a net chirality. Therefore, one can imagine terms in the edge effective
theory that generate a gap. To construct this edge, let us first
consider the situation in Fig.~\ref{fig:edges}(b), in which we have
all three states, $X$, $M$, and $SC$:
a thin strip of $M$ separates $X$ from $SC$. The edge between $X$ and
$M$ has been discussed above. The edge between $M$ and $SC$
has a gapless Majorana fermion excitation.
A real 2D $SC$ state will have a gapless charged mode and, therefore,
charged excitations at the edge can leak into the bulk.
However, we can consider, as a matter of principle, a fixed
non-dynamical superconducting order parameter $\Delta$
which renders the surface of a 3D TI gapped. Such a system
is fully-gapped. Now, as we make the $M$ strip narrower, the two edges
will become coupled and can be treated as a single edge
between $X$ and $SC$. This edge has
two right-moving Majorana fermions, one from the
$X$-$M$ boundary and one from the $M$-$SC$ boundary.
There is also the charged boson of the $X$-$M$ boundary, which is left-moving, so that
$c=1$, $\overline{c}=1$. The action takes the form
\begin{multline}
S = \int dt \, dx \, \Bigl[i \psi_1({\partial_t}+{v_1}\partial_x)\psi_1 +
 i \psi_2({\partial_t}+{v_1}\partial_x)\psi_2\\
+ \frac{2}{4\pi} \partial_x \phi_2 (-\partial_t + v \partial_x)\phi_2
+ \lambda \cos(2{\phi_2}) i\psi_1 \psi_2 \Bigr]
\end{multline}
The final term is the leading coupling between the
$X$-$M$ boundary and the $M$-$SC$ boundary. It
transfers an electron from one to the other.
Let us consider, for simplicity, the case
${v_1}={v_2}={v_f}$. Then we can bosonize the two Majorana fermions
by introducing a new bosonic field $\phi_1$ according to
$e^{i\phi_1} = \psi_1 + i\psi_2$. The action now takes the form:
\begin{multline}
S = \int dt \, dx \, \Bigl[\frac{1}{4\pi} \partial_x \phi_1 (\partial_t + {v_f} \partial_x)\phi_1\\
+ \frac{2}{4\pi} \partial_x \phi_2 (-\partial_t + {v} \partial_x)\phi_2
+ \frac{1}{2\pi} \lambda \cos(2{\phi_2})\, \partial_x \phi_1 \Bigr]
\end{multline}
The dimension-$1$ operators ${J^x}=\cos(2{\phi_2})$,
${J^y}=\sin(2{\phi_2})$, ${J^z}=\frac{1}{2\pi}\partial_x\phi_2$
form an $SU(2)_1$ current algebra. Consequently, we can perform
an $SU(2)$ rotation by $\pi/2$ about the $J^y$ axis to transform the
action to
\begin{multline}
S = \int dt \, dx \, \Bigl[\frac{1}{4\pi} \partial_x \phi_1 (\partial_t + {v_f} \partial_x)\phi_1\\
+ \frac{2}{4\pi} \partial_x \phi_2 (-\partial_t + {v} \partial_x)\phi_2
+ \frac{1}{(2\pi)^2} \lambda \partial_x\phi_2\, \partial_x \phi_1 \Bigr]
\label{eqn:X-SC-edge}
\end{multline}
This is essentially the edge theory of an Abelian $(1,1,3)$ quantum Hall
state~\cite{Halperin83}, which is the same as that of the
$(3,3,1)$ state, except with the two modes moving in opposite
directions, rather than the same direction. However, the theory
in Eq.~(\ref{eqn:X-SC-edge}) differs
from the edge theory of the $(1,1,3)$ quantum Hall state in
that only anti-periodic boundary conditions are allowed
for the fermions (since the superconducting order parameter is
non-dynamical, there are no vortices in the SC). Therefore, the only
allowed operators in the theory are:
$1, e^{i\phi_2}, e^{i({\phi_1}-{\phi_2})}, e^{i\phi_1}$. For $\lambda=0$,
they have right and left scaling dimensions equal to, respectively,
$(0,0), (1/4,0), (1/4,1/2), (0,1/2)$. Therefore, this is a theory
with the vacuum, $\theta=\pm i$ semions, and a fermion.
However, for generic $\lambda$, including $\lambda=0$,
it is not invariant under the time-reversal transformation
${\phi_1}\rightarrow {\phi_1}-2{\phi_2}$, ${\phi_2}\rightarrow {\phi_1}-{\phi_2}$.
This is because, by introducing a strip of $M$ between $X$ and $SC$,
we have broken time-reversal symmetry. When shrinking this
strip, we must be careful to do so in a way that preserves time-reversal
symmetry. This can be done if we demand that $v={v_f}=-\lambda/2$,
in which case Eq.~(\ref{eqn:X-SC-edge}) is invariant under
${\phi_1}\rightarrow {\phi_1}-2{\phi_2}$, ${\phi_2}\rightarrow {\phi_1}-{\phi_2}$,
which maps the $\theta=\pm i$ semion into the $\theta=\mp i$ semion.
For these values of couplings, $e^{i\phi_2}$ and
$e^{i({\phi_1}-{\phi_2})}$ have scaling dimensions
$(\frac{1}{8}(\sqrt{2}+1), \frac{1}{8}(\sqrt{2}-1))$ and
$(\frac{1}{8}(\sqrt{2}-1), \frac{1}{8}(\sqrt{2}+1))$, respectively.

\subsection{Gauge invariance and $\theta_{(2,2)}$}

The conclusion $\theta_{(2,2)}=\pm i$ is consistent with gauge invariance~\cite{Kivelson85}.
Suppose that, in the geometry in Fig.~\ref{fig:slab}(a), there is
an $(2,2)$ quasiparticle in the $X$ region, i.e. in the interior of the annulus
on the top surface bounded by the edges $E_1$ and $E_2$.
If there are no other quasiparticles in the annulus (either in its bulk or at its
edge) and there is no magnetic flux threaded through the hole,
then when we take this quasiparticle around the annulus, no phase results
(apart from the non-topological dynamical phase that depends on the details
of the motion).
Suppose we adiabatically increase the flux through the ring until it is $\Phi_0$,
thereby creating the quasiparticle $(2,2)$ at the inner edge $E_1$.
The phase obtained by taking the bulk $(2,2)$ quasiparticle around the annulus
is $e^{i \phi} = e^{i\pi} \left[R^{(2,2) (2,2)}_{(4,4)}\right]^2$. The first factor,
$e^{i\pi}$ is the electromagnetic Aharonov-Bohm phase for a charge $e/2$
going around flux $\Phi_0$, while the second factor is the Abelian braiding phase.
However, all physical properties must be periodic in the flux with period $\Phi_0$,
so $e^{i \phi}=1$. It follows that $R^{(2,2) (2,2)}_{(4,4)}=\pm i$, which, using Eq.~(\ref{eq:twist}), implies that
$\theta_{(2,2)}=\pm i$.

\subsection{The electron is the $\psi_4 = (4,0)$ quasiparticle}

In our arguments, we assumed that the $(4,0)$ quasiparticle type is assigned to the electron.
We now give an argument why this is the case.
Consider a TI in a solid torus shape, with the top half surface covered with $M_{+}$
and the bottom half covered with $M_{-}$.
There will be two circular boundaries between the $M_{+}$ and $M_{-}$ regions.
Now, split one of those boundaries by inserting a thin annular strip of $X$,
as in Fig.~\ref{fig:edges}(a). Insert flux $\Phi_0$ through the torus. This transfers charge $e$
to the boundary between $M_{+}$ and $M_{-}$. This must be an electron since this edge
is just an IQH edge. Flux insertion also puts charge $-e/2$ at the $X$-$M_{+}$
boundary and charge $-e/2$ at the $X$-$M_{-}$ boundary. The two possibilities
are $I_2$ and $\psi_2$. Consequently, the quasiparticle at the $X$-$M_{+}$
boundary and the quasiparticle at the $X$-$M_{-}$ boundary fuse to
either $I_4$ or $\psi_4$. Since the $M_{+}$-$M_{-}$ boundary
has an electron, we must identify this quasiparticle with an electron,
which means that it must be a charge-$e$ fermion. Consequently
this quasiparticle is $\psi_4$, and we must identify it with an electron.

\subsection{Topological twist factor of $\sigma$ quasiparticles}

The topological phase $X$ contains $\sigma_{2j+1}$ quasiparticles with twist factors $\theta_{\sigma_{2j+1}} = \pm 1$,
which we deduced from time-reversal and charge conservation.
This result is, at first glance, very natural if we think
about a continuous phase transition from the $SC$ phase into $X$.

\begin{figure}
\centerline{
\includegraphics[width=4in]{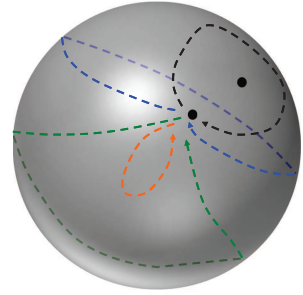}
}
\caption{When a ${\sigma_1}$ quasiparticle is taken around a
${\sigma_7}$ quasiparticle with which it fuses to $I_0$,
the result is $R^{\sigma_1 \sigma_7}_{I_0}R^{\sigma_7 \sigma_1}_{I_0}$ when this is done
along the black trajectory. When the trajectory is deformed to the blue
one, then to the green one and, finally, to the orange one, the resulting phase
is zero if the quasiparticle does not couple to the curvature.}
\label{fig:sphere-braid}
\end{figure}

In an ordinary 2D $p_x + i p_y$ superconductor, the condensate breaks rotational symmetry, while leaving a combination of rotational and gauge symmetries
preserved. Consequently, the $p_x + i p_y$ vortices are coupled to the spatial curvature. By contrast, since the $SC$ phase on the surface of a 3D TI has $s$-wave symmetry, the superconducting order parameter does not couple to the
curvature of the surface, because the condensate
does not break rotational symmetry. Therefore, the vortices are
also not coupled to spatial curvature. The absence of coupling
to the curvature implies that $\theta = \pm 1$. This means, for instance,
that the phase for taking one quasiparticle around another must be $0$ when
the two quasiparticles fuse to the identity, because such a trajectory can
be continuously deformed into a trivial one on the sphere (as depicted
in Fig.~\ref{fig:sphere-braid}) and there is
no correction due to the curvature of the sphere.

If we assume that the condensation of $8\pi$ vortices does not change the
absence of curvature coupling, then we expect $\theta=\pm 1$ for quasiparticles in $X$.
This is, in fact, correct for $\sigma_{2j+1}$ quasiparticles:
$\theta_{\sigma_1}=\theta_{\sigma_7}=-\theta_{\sigma_3}=-\theta_{\sigma_5} =\pm 1$.
However, this does not hold for the Abelian quasiparticles, some of which
have $\theta=\pm i$. Thus, $8\pi$ vortex condensation affects the topological
twist factors of some of the Abelian vortices. It would be interesting to
understand the mechanism for this.

\section{Conclusion}

Most of the time-reversal invariant topological phases studied
thus far have been doubled Chern-Simons theories, in which there
are essentially two opposite-chirality copies of the system, and discrete gauge theories (quantum doubles D($G$) of a group $G$).
Time-reversal acts in a relatively straightforward manner for these theories.
In this paper, we have seen
a rather different sort of time-reversal invariant topological phase.
It can be understood in terms of two opposite chirality sectors, but
they are very different (although only a restricted subset of their product is allowed):
one sector is electrically neutral and non-Abelian while
the other is electrically charged and Abelian. Nevertheless, under the unconventional
time-reversal transformation discussed above, the state is time-reversal invariant.
It would be interesting to see if there are more direct physical consequences
of this unusual transformation law and also whether there are
novel time-reversal invariant versions of other topological phases,
e.g. one that contains Fibonacci anyons.

Although this paper is focussed primarily on a question of principle,
we now comment briefly on possible physical realizations. The surface
of a topological insulator such as Bi$_2$Se$_3$ can be gapped by
covering it with a superconducting film. However, it would then be necessary
to destroy the superconducting order so that the surface can enter
a topological phase. It is not clear how to do this. One could imagine patterning
the superconducting film into a Josephson junction array and then
increasing the charging energy for each superconducting island until
superconductivity is destroyed, but it is possible that this will leave
gapless regions between the superconducting islands. An equally serious
problem is that all putative topological insulators are not truly insulating
and, in fact, have relatively high bulk carrier densities. The topological
properties described here would require true insulating behavior in
the 3D bulk in order to be observed, but there may be a range of time
scales over which they are observable for sufficiently-low bulk conductivity.
Alternatively, it may be possible to realize such a phase in a system of cold atoms,
where it has been proposed~\cite{Anderson12} that a synthetic 3D spin-orbit coupling can be realized, and there are proposed realizations
of 3D topological insulators~\cite{Mazza12}.

During completion of this manuscript, we learned of work by M. Metlitski {\it et al.}~\cite{Metlitski-unpublished} and C. Wang {\it et al.}~\cite{Wang-unpublished} constructing another time-reversal
invariant and charge-conserving topological phase, described by the anyon model $\left. \text{Ising} \times \mathbb{Z}_8^{(1/2)} \right|_{\mathcal{C}} \times \mathbb{Z}_2^{(-1/2)}$ (i.e. the Moore-Read theory times a $\theta=-i$ semion theory) which has 24 quasiparticle types,
but is more directly connected to the superconducting state of the surface of a fermionic TI.
We also learned of work by X. Chen {\it et al.}~\cite{Chen-unpublished} discussing both of these
phases. The relation between these two phases is presently unclear,
but several possibilities were delineated by X. Chen {\it et al}.

\section{Acknowledgements}
We would like to thank W.~Bishara, X.~Chen, L.~Fidkowski, M.~Fisher, T.~Grover,
M.~Metlistski, J.~Slingerland, A.~Vishwanath, and Z.~Wang for discussions.
We thank the Aspen Center for Physics for hospitality and
support under the NSF Grant No. 1066293.
C.~N. has been partially supported by the DARPA QuEST
program and AFOSR under grant FA9550-10-1- 0524. X.~L.~Q. is supported by the National Science Foundation through the grant No. DMR-1151786.

\appendix

\section{Anyon Models}
\label{sec:anyon_models}

In this appendix, we briefly review the basic fusion and braiding properties of quasiparticles in $(2+1)$D systems, as described by anyon models (a.k.a. unitary braided tensor categories). For additional details, see Refs.~\onlinecite{Kitaev06a,Bonderson07b} and references therein.

An anyon model is defined by a finite set $\mathcal{C}$ of topological charges, which obey a commutative, associative fusion algebra
\begin{equation}
a \times b = \sum\limits_{c \in \mathcal{C}} N_{ab}^{c} \, c
\end{equation}
where $N_{ab}^{c}$ are positive integers indicating the number of distinct ways charges $a$ and $b$ can be combined to produce charge $c$. There is a unique ``vacuum'' charge, denoted $0$ or $I$, which has trivial fusion (and braiding) with all other charges (for example $N_{a 0}^{c} = \delta_{ac}$) and which defines the unique conjugate $\bar{a}$ of each topological charge $a$ via $N_{a b}^{0} = \delta_{\bar{a} b}$.

Each fusion product has an associated vector space $V_{ab}^{c}$ with $\dim{V_{ab}^{c}} = N_{ab}^{c}$, and its dual (splitting) space $V^{ab}_{c}$. The states in these fusion and splitting spaces are assigned to trivalent vertices with the appropriately corresponding topological charges:
\begin{equation}
\pspicture[shift=-0.6](-0.1,-0.2)(1.5,-1.2)
  \small
  \psset{linewidth=0.9pt,linecolor=black,arrowscale=1.5,arrowinset=0.15}
  \psline(0.7,0)(0.7,-0.55)
  \psline(0.7,-0.55) (0.25,-1)
  \psline(0.7,-0.55) (1.15,-1)	
  \rput[tl]{0}(0.4,0){$c$}
  \rput[br]{0}(1.4,-0.95){$b$}
  \rput[bl]{0}(0,-0.95){$a$}
 \scriptsize
  \rput[bl]{0}(0.85,-0.5){$\mu$}
  \endpspicture
=\left\langle a,b;c,\mu \right| \in
V_{ab}^{c} ,
\label{eq:bra}
\end{equation}
\begin{equation}
\pspicture[shift=-0.6](-0.1,-0.2)(1.5,1.2)
  \small
  \psset{linewidth=0.9pt,linecolor=black,arrowscale=1.5,arrowinset=0.15}
  \psline(0.7,0)(0.7,0.55)
  \psline(0.7,0.55) (0.25,1)
  \psline(0.7,0.55) (1.15,1)	
  \rput[bl]{0}(0.4,0){$c$}
  \rput[br]{0}(1.4,0.8){$b$}
  \rput[bl]{0}(0,0.8){$a$}
 \scriptsize
  \rput[bl]{0}(0.85,0.35){$\mu$}
  \endpspicture
=\left| a,b;c,\mu \right\rangle \in
V_{c}^{ab},
\label{eq:ket}
\end{equation}
where $\mu=1,\ldots ,N_{ab}^{c}$. Most anyon models of interest have no fusion multiplicities, i.e. $N_{ab}^{c}=0$ or $1$, in which case the vertex labels $\mu$ are trivial and can be left implicit (as is done throughout this paper).
General states and operators are described using fusion/splitting trees constructed by connecting lines with the same topological charge.

Associativity of fusion is represented in the state space by the $F$-symbols, which (similar to $6j$-symbols) provide a unitary isomorphism relating states written in different bases distinguished by the order of fusion. Diagrammatically, this is represented as
\begin{equation}
\label{eq:F_move}
  \pspicture[shift=-1.0](0,-0.45)(1.8,1.8)
  \small
  \psset{linewidth=0.9pt,linecolor=black,arrowscale=1.5,arrowinset=0.15}
  \psline(0.2,1.5)(1,0.5)
  \psline(1,0.5)(1,0)
  \psline(1.8,1.5) (1,0.5)
  \psline(0.6,1) (1,1.5)
   \rput[bl]{0}(0.05,1.6){$a$}
   \rput[bl]{0}(0.95,1.6){$b$}
   \rput[bl]{0}(1.75,1.6){${c}$}
   \rput[bl]{0}(0.5,0.5){$e$}
   \rput[bl]{0}(0.9,-0.3){$d$}
  \endpspicture
= \sum_{f} \left[F_d^{abc}\right]_{ef}
 \pspicture[shift=-1.0](0,-0.45)(1.8,1.8)
  \small
  \psset{linewidth=0.9pt,linecolor=black,arrowscale=1.5,arrowinset=0.15}
  \psline(0.2,1.5)(1,0.5)
  \psline(1,0.5)(1,0)
  \psline(1.8,1.5) (1,0.5)
  \psline(1.4,1) (1,1.5)
   \rput[bl]{0}(0.05,1.6){$a$}
   \rput[bl]{0}(0.95,1.6){$b$}
   \rput[bl]{0}(1.75,1.6){${c}$}
   \rput[bl]{0}(1.25,0.45){$f$}
   \rput[bl]{0}(0.9,-0.3){$d$}
  \endpspicture
.
\end{equation}

The counterclockwise braiding exchange operator of topological charges $a$ and $b$ is represented diagrammatically as
\begin{equation}
\label{eq:braid}
R_{ab}=
\pspicture[shift=-0.6](-0.1,-0.2)(1.3,1.05)
\small
  \psset{linewidth=0.9pt,linecolor=black,arrowscale=1.5,arrowinset=0.15}
  \psline(0.96,0.05)(0.2,1)
  \psline(0.24,0.05)(1,1)
  \psline[border=2pt](0.24,0.05)(0.92,0.9)
  \rput[bl]{0}(-0.02,0){$a$}
  \rput[br]{0}(1.2,0){$b$}
  \endpspicture
.
\end{equation}
The action of this operator on the state space can be described by the $R$-symbols, which represent the unitary operator for exchanging two anyons in a specific fusion channel, and are obtained by applying the exchange operator to the corresponding trivalent vertices
\begin{equation}
\label{eq:R_move}
\pspicture[shift=-0.65](-0.1,-0.2)(1.5,1.2)
  \small
  \psset{linewidth=0.9pt,linecolor=black,arrowscale=1.5,arrowinset=0.15}
  \psline(0.7,0)(0.7,0.5)
  \psarc(0.8,0.6732051){0.2}{120}{240}
  \psarc(0.6,0.6732051){0.2}{-60}{35}
  \psline (0.6134,0.896410)(0.267,1.09641)
  \psline(0.7,0.846410) (1.1330,1.096410)	
  \rput[bl]{0}(0.4,0){$c$}
  \rput[br]{0}(1.35,0.85){$a$}
  \rput[bl]{0}(0.05,0.85){$b$}
  \endpspicture
= R_{c}^{ab}
\pspicture[shift=-0.65](-0.1,-0.2)(1.5,1.2)
  \small
  \psset{linewidth=0.9pt,linecolor=black,arrowscale=1.5,arrowinset=0.15}
  \psline(0.7,0)(0.7,0.55)
  \psline(0.7,0.55) (0.25,1)
  \psline(0.7,0.55) (1.15,1)	
  \rput[bl]{0}(0.4,0){$c$}
  \rput[br]{0}(1.4,0.8){$a$}
  \rput[bl]{0}(0,0.8){$b$}
  \endpspicture
.
\end{equation}

An anyon model is defined entirely by its $N_{ab}^{c}$, $F$-symbols, and $R$-symbols. The $N_{ab}^{c}$ must provide an associative and commutative algebra. The $F$-symbols and $R$-symbols are constrained by the ``coherence conditions'' (also known as the ``polynomial equations''), which ensure that any two series of $F$ and/or $R$ transformations are equivalent if they start in the same state space and end in the same state space~\cite{MacLane63}. Physically, these consistency conditions are interpreted as enforcing locality in fusion and braiding processes.

Distinct sets of $F$-symbols and $R$-symbols describe equivalent theories if they can be related by a gauge transformation given by unitary transformations acting on the fusion/splitting state spaces $V_{c}^{ab}$ and $V^{c}_{ab}$, which can be though of as a simple redefinition of the basis states as
\begin{equation}
\label{eq:gauge}
\widetilde{ \left| a,b;c,\mu \right\rangle } = \sum_{\nu} \left[u^{ab}_{c}\right]_{\mu \nu} \left| a,b;c,\nu \right\rangle
\end{equation}
where $u^{ab}_{c}$ is the unitary transformation, equal to a phase factor when $N_{ab}^{c}=1$. Such gauge transformations modify the $F$-symbols and $R$-symbols as
\begin{eqnarray}
\left[\widetilde{F}_d^{abc}\right]_{ef} &=& \frac{ u^{ab}_{e} u^{ec}_{d} }{u^{bc}_{f} u^{af}_{d}} \left[F_d^{abc}\right]_{ef} \\
\widetilde{R}_{c}^{ab} &=& \frac{ u^{ab}_{c} }{u^{ba}_{c}}  R_{c}^{ab}
.
\end{eqnarray}
(One must be careful not to use the gauge freedom associated with $u^{a 0}_{a}$ and $u^{0 b}_{b}$ to ensure that fusion and braiding with the vacuum $0$ is trivial.)
It is often useful to consider quantities of the anyon model that are invariant under such gauge transformation. The most relevant gauge invariant quantities are the quantum dimensions $d_a$ and topological twist factors $\theta_a$, since these, together with the fusion coefficients $N_{ab}^{c}$, usually uniquely specify the theory (there are no known counterexamples).

The quantum dimension of topological charge $a$
\begin{equation}
\label{eq:q_dim}
d_{a} = d_{\bar{a}} \equiv \left| \left[F_a^{a \bar{a} a}\right]_{00} \right|^{-1}
\end{equation}
is also equal to the largest eigenvalue of the matrix $M^{(a)}$ defined by $[M^{(a)}]_{bc} \equiv N^{c}_{ab}$, and so describes how the dimensionality of the state space grows asymptotically as one introduces more quasiparticles of charge $a$ (i.e. dim$[V^{a \ldots a}]\sim d_a ^{n}$ when the number $n$ of charge $a$ quasiparticles is large). The quantum dimensions must satisfy the property
\begin{equation}
\label{eq:q_dim_relation}
d_a d_b = \sum_{c} N^{c}_{ab} d_c .
\end{equation}

The topological twist factor of topological charge $a$
\begin{equation}
\label{eq:twist}
\theta_{a} = \theta_{\bar{a}} =\sum_{c\in \mathcal{C}} \frac{d_c}{d_a} R^{aa}_{c} = d_a \left[F_a^{a \bar{a} a}\right]_{00} \left(R^{\bar{a} a}_{0}\right)^{\ast}
\end{equation}
are roots of unity associated with the braiding statistics of the anyons of corresponding topological charge. These are the phases that would arise if one were able to rotate a quasiparticle carrying topological charge $a$, or if one had a quasiparticle of charge $a$, created an $a$-$\bar{a}$ pair from vacuum, exchanged the two $a$ quasiparticles in a counterclockwise sense, and then annihilated (to vacuum) the original $a$ quasiparticle with the $\bar{a}$ quasiparticle. This can be represented diagrammatically as
\begin{equation}
\pspicture[shift=-1.0](-0.5,-1.1)(2.6,1.3)
  \small
   \rput[bl]{0}(1.4,1.1){$a$}
  \psbezier[linewidth=0.9pt,linecolor=black,border=0.1](1.5,1.0)(1.5,-0.5)(2.5,-0.5)(2.5,0.0)
  \psbezier[linewidth=0.9pt,linecolor=black,border=0.1](1.5,-1.0)(1.5,0.5)(2.5,0.5)(2.5,0.0)
\endpspicture
= \theta_{a}
\pspicture[shift=-1.0](-0.2,-1.1)(0.5,1.3)
  \small
  \psset{linewidth=0.9pt,linecolor=black,arrowscale=1.5,arrowinset=0.15}
  \psline(0.0,1.0)(0.0,-1.0)
   \rput[bl]{0}(-0.1,1.1){$a$}
\endpspicture
.
\end{equation}
This can be used to show that the $R$-symbols satisfy the ``ribbon property''
\begin{equation}
\label{eq:ribbon}
R^{ba}_{c} R^{ab}_{c} = \frac{\theta_{c}}{\theta_{a} \theta_{b}}
.
\end{equation}
We note that, for an Abelian anyon $a$, Eq.~(\ref{eq:twist}) gives $\theta_{a} = R^{a a}_{a^2}$ (where $a^2 = a \times a$) and thus, using Eq.~(\ref{eq:ribbon}), we have
\begin{equation}
\label{eq:Abelian_twists}
\theta_{a^2} = \theta_{a}^4 \qquad \text{ for $a$ Abelian}
.
\end{equation}
The twist factors can be used to define the topological $T$-matrix by $T_{ab} = \theta_a \delta_{ab}$.

Another significant invariant is the topological $S$-matrix
\begin{equation}
S_{ab} = \frac{1}{\mathcal{D}} \sum_{c\in \mathcal{C}} N_{ab}^{c} d_c \frac{  \theta_c }{ \theta_a \theta_b}
\end{equation}
where $\mathcal{D} = \sqrt{\sum_{a\in \mathcal{C}} d_a^2}$ is the total quantum dimension.

\subsection{Modularity}
\label{sec:modularity_anyon}

An anyon model is ``modular'' and associated with a TQFT iff $S$ is unitary. In this case, the $S$-matrix together with the $T$-matrix, $T_{ab} = \theta_a \delta_{ab}$, and the charge conjugation matrix $C_{ab} =\delta_{\bar{a} b}$ obey the modular relations
\begin{equation}
(ST)^3 = \Theta C, \quad S^2 = C, \quad C^2 = 1
\end{equation}
where
\begin{equation}
\Theta = \frac{1}{\mathcal{D}} \sum_{a\in \mathcal{C}} d_a^2 \theta_a = e^{i \frac{2\pi}{8} c_{-}}
\end{equation}
is a root of unity and $c_{-} \equiv c - \bar{c}$ is the chiral central charge. These correspond to the TQFT's representation of the respective modular transformations.

If an anyon model satisfies the following conditions:
\begin{enumerate}
\item There is a fermion with topological charge $z$ (i.e. $z \times z = I$ and $\theta_z=-1$) such that $a \times z \neq a$ and $\theta_{a \times z} = - \theta_{a}$ for all topological charges $a$ (which implies that $z$ is mutually local with all topological charges. It follows from these conditions that the topological $S$-matrix and $T$-matrix take the form
\begin{eqnarray}
\label{eq:Z_2_S}
S &=& S_z \otimes \widehat{S} , \\
T &=& T_z \otimes \widehat{T},
\end{eqnarray}
where
\begin{eqnarray}
S_z &=& \frac{1}{\sqrt{2}} \left[
\begin{array}{cc}
1 & 1 \\
1 & 1
\end{array}
\right] ,\\
T_z &=& \left[
\begin{array}{cc}
1 & 0 \\
0 & -1
\end{array}
\right]
\end{eqnarray}
are the $S$-matrix and $T$-matrix of the $\mathbb{Z}_{2}^{(1)}$ subsector generated by $z$.

\item The $\widehat{S}$ in Eq.~(\ref{eq:Z_2_S}) is unitary.
\end{enumerate}
then it is not modular, since the $S$-matrix is degenerate, but the notion of modularity can be partially salvaged by applying a $\mathbb{Z}_2$-grading. In particular, one forms $\mathbb{Z}_2$ doublets of topological charge $\widehat{a} = \{a , a \times z\}$, so that one can use the unitary $\widehat{S}$ to represent the corresponding modular transformation. In this way, the anyon model is associated with a $\mathbb{Z}_2$-graded TQFT (or spin field theory)~\cite{Dijkgraaf90}. These $\mathbb{Z}_2$-graded theories are only defined on manifolds that permit a spin structure and the Hilbert space for a given manifold decomposes into a direct sum over the spin structure. Thus, in order to obtain maps between states with the same spin structure, one must restrict to modular transformations that preserve the spin structure. We note that $\widehat{T}$ is not unambiguously defined by the above expression, since $\theta_{a \times z} = - \theta_{a}$, but $\widehat{T}^2$ is unambiguous.

\subsection{Time reversal}
\label{sec:time_reversal_anyon}

The action of the (anti-unitary and anti-linear) time reversal transformation $\mathcal{T}$ on anyon models can potentially map topological charge values to different values
\begin{equation}
\mathcal{T} : a \mapsto a'
\end{equation}
and hence
\begin{equation}
\mathcal{T} : N_{ab}^{c} \mapsto N_{a' b'}^{c'}
\end{equation}

The action on $F$-symbols and $R$-symbols maps them to their inverses. Since we assume the anyon models are unitary, this means
\begin{eqnarray}
&& \mathcal{T} : \left[F^{a b c}_{d}\right]_{ef} \mapsto \left[F^{a' b' c'}_{d'}\right]_{e' f'}^{\ast} \\
&& \mathcal{T} : R^{a b}_{c} \mapsto \left[ R^{a' b'}_{c'}\right]^{\ast}
.
\end{eqnarray}
It follows that
\begin{eqnarray}
&& \mathcal{T} : d_a \mapsto d_{a'} \\
&& \mathcal{T} : \theta_{a} \mapsto \theta_{a'}^{\ast}
.
\end{eqnarray}

It is sometimes useful to incorporate a gauge transformation $U$, as in Eq.~(\ref{eq:gauge}), in the definition of time reversal, i.e. $\widetilde{\mathcal{T}} = U \mathcal{T}$. In this case, the action on gauge invariant quantities is unchanged, while the action on state vectors, $F$-symbols, and $R$-symbols becomes
\begin{eqnarray}
&& \widetilde{ \mathcal{T} } :  \left| a,b;c \right\rangle \mapsto \widetilde{ \left| a',b';c' \right\rangle } = u^{a' b'}_{c'} \left| a',b';c' \right\rangle \\
&& \widetilde{ \mathcal{T} } : \left[F^{a b c}_{d}\right]_{ef} \mapsto \frac{ u^{a' b'}_{e'} u^{e' c'}_{d'} }{u^{b' c'}_{f'} u^{a' f'}_{d'}} \left[F^{a' b' c'}_{d'}\right]_{e' f'}^{\ast} \\
&& \widetilde{ \mathcal{T} } : R^{a b}_{c} \mapsto \frac{ u^{a' b'}_{c} }{u^{b' a'}_{c'} } \left[ R^{a' b'}_{c'}\right]^{\ast}
.
\end{eqnarray}

\subsection{Ising$^{(n)}$ anyon models}
\label{sec:Ising_anyons}

The Ising anyon model has topological charge types $I$, $\sigma$, and $\psi$.
The fusion algebra, specified in Eq.~(\ref{eq:Ising_fusion_rules}), indicates that the non-zero fusion coefficients are
\begin{equation}
\label{eq:Ising_N}
N_{I A}^{A} = N_{A I}^{A} = N_{\psi \psi}^{I} = N_{\sigma \psi}^{\sigma} = N_{\psi \sigma}^{\sigma} = N_{\sigma \sigma}^{I}= N_{\sigma \sigma}^{\psi}=1
,
\end{equation}
where $A = I$, $\sigma$, and $\psi$.

The non-trivial $F$-symbols are
\begin{eqnarray}
F^{\psi \sigma \psi}_{\sigma} &=& F^{\sigma \psi \sigma}_{\psi} =-1 \\
\left[ F^{\sigma \sigma \sigma}_{\sigma} \right]_{EF} &=&
\frac{1}{\sqrt{2}} \left[
\begin{matrix}
1 & 1\cr 1 & -1
\end{matrix}
\right]_{EF}
\end{eqnarray}
where the column and row values $E$ and $F$ of the matrix take values $I$ and $\psi$ (in order), and the rest are either equal to $1$, if the corresponding fusion trees are allowed by the fusion rules, or equal to $0$ if they are not allowed by fusion.

The $R$-symbols are
\begin{eqnarray}
R^{I A}_{A} &=& R^{A I}_{A} = 1 \notag \\
R^{\psi \psi }_{I} &=& - 1 \notag \\
R^{\psi \sigma}_{\sigma} &=& R^{\sigma \psi}_{\sigma} = -i \notag \\
R^{\sigma \sigma}_{I} &=& e^{-i \frac{\pi}{8}} \notag \\
R^{\sigma \sigma}_{\psi} &=& e^{i \frac{3 \pi}{8}}
.
\end{eqnarray}
(The $R$-symbols for fusion vertices not allowed by the fusion rules are set to $0$.)

The quantum dimensions are $d_{I} = d_{\psi} =1$ and $d_{\sigma} =\sqrt{2}$. The twist factors are $\theta_I=1$, $\theta_\sigma=e^{i \frac{\pi}{8}}$, and $\theta_\psi=-1$. The $S$-matrix is
\begin{equation}
S = \frac{1}{2} \left[
\begin{matrix}
1 & \sqrt{2} & 1 \cr
\sqrt{2} & 0 & -\sqrt{2} \cr
1 & -\sqrt{2} & 1
\end{matrix}
\right]
.
\end{equation}

It is worth mentioning that the Ising anyon model is closely related to seven other anyon models that are its ``Galois conjugates,'' which we will denote as Ising$^{(n)}$ where $n\in \mathbb{Z}$ is only distinguished modulo $8$ (with $n=0$ being the Ising anyon model). In simpler terms, these anyon models have the same topological charges and fusion algebra as Ising, but slightly different $F$-symbols and $R$-symbols, which can be obtained from the following modifications
\begin{eqnarray}
F^{\sigma \sigma \sigma}_{\sigma} &=&
\frac{\varkappa_{\sigma}}{\sqrt{2}} \left[
\begin{matrix}
1 & 1\cr 1 & -1
\end{matrix}
\right] \notag \\
R^{\psi \sigma}_{\sigma} &=& R^{\sigma \psi}_{\sigma} = (-i)^{2n+1} \notag \\
R^{\sigma \sigma}_{I} &=& \varkappa_{\sigma} e^{-i \frac{\pi (2n+1)}{8}} \notag \\
R^{\sigma \sigma}_{\psi} &=& \varkappa_{\sigma} e^{i \frac{3 \pi (2n+1)}{8}}
,
\end{eqnarray}
where $\varkappa_{\sigma}= (-1)^{n(n+1)/2}$. The twist factor $\theta_\sigma=e^{i \frac{\pi (2n+1)}{8}}$ uniquely distinguishes the eight Ising$^{(n)}$ anyon models.

We note that these eight anyon models are the only gauge-inequivalent anyon models with the Ising fusion algebra that satisfy the coherence conditions, they are all unitary (which is not generally the case with Galois conjugates), and they are all modular.

\subsection{$\mathbb{Z}^{(w)}_{N}$ anyon models}

The $\mathbb{Z}^{(w)}_{N}$ anyon models for $N$ a positive integer can have $w\in \mathbb{Z}$ for all $N$, and also $w\in \mathbb{Z}+\frac{1}{2}$ for $N$ even. The topological charges $0,1,\ldots, N-1$ obey the fusion algebra $j \times k = [j+k]_{N}$, where $[j]_{N} = j (\text{mod } N)$, giving the fusion coefficients
\begin{equation}
N_{j k}^{l} = \delta_{l,[j+k]_N }
.
\end{equation}

The nonzero $F$-symbols are
\begin{equation}
\left[F^{j k l}_{[j+k+l]_N } \right]_{[j+k]_N [k+l]_N } = e^{i \frac{2 \pi w}{N} j (k + l - [k+l]_{N} )}
\end{equation}
(the rest not being allowed by fusion). Note that these are all equal to $1$ when $w \in \mathbb{Z}$, while some of them equal $-1$ when $w\in \mathbb{Z}+\frac{1}{2}$ and these minus signs cannot simply be gauged away.

The nonzero $R$-symbols are
\begin{equation}
R^{j k }_{[j+k]_N } = e^{i \frac{2 \pi w}{N} j k }
.
\end{equation}

The quantum dimensions are $d_{j} =1$, the twist factors are $\theta_j = e^{i \frac{2 \pi w}{N} j^2 }$, and the $S$-matrix has elements $S_{jk} = \frac{1}{\sqrt{N} } e^{i \frac{4 \pi w}{N} j k }$.

It should be clear that $\mathbb{Z}^{(w)}_{N}=\mathbb{Z}^{(w')}_{N}$ if $w = w' (\text{mod }N)$. However, it is sometimes possible to further equate these anyon models when $w \neq w' (\text{mod }N)$ by relabeling the topological charges, e.g. $\mathbb{Z}^{(1)}_{5} = \mathbb{Z}^{(4)}_{5}$ by letting $0=0'$, $2=1'$, $4=2'$, $1=3'$, and $3=4'$.

\subsection{Minimal anyon model describing the phase $X$}
\label{sec:X_min_anyon_model}

We now consider anyon models with topological charges $\mathcal{C} = \left\{ I_{2j}, \psi_{2j}, \sigma_{2j+1} | j =0,1,2,3 \right\} $  ($I_0$ is the vacuum/trivial topological charge) and corresponding fusion algebra
\begin{eqnarray}
I_{2j} \times I_{2k} &=& I_{[2j+2k]_8}, \notag \\
I_{2j} \times \psi_{2k} &=& \psi_{2k} \times I_{2j} = \psi_{[2j+2k]_8}, \notag \\
I_{2j} \times \sigma_{2k+1} &=& \sigma_{2k+1} \times I_{2j} = \sigma_{[2j+2k+1]_8}, \notag \\
\psi_{2j} \times \psi_{2k} &=& I_{[2j+2k]_8}, \notag \\
\psi_{2j} \times \sigma_{2k+1} &=& \sigma_{2k+1} \times \psi_{2j} = \sigma_{[2j+2k+1]_8}, \notag \\
\sigma_{2j+1} \times \sigma_{2k-1} &=& I_{[2j+2k]_8}+ \psi_{[2j+2k]_8},
\end{eqnarray}
corresponding to the restricted product of fusion algebras $ \left. \text{Ising} \times \mathbb{Z}_8 \right|_{\mathcal{C}} $, using the equivalence with the shorthand notation $A_j \equiv (A,j)$ with $A \in \mathcal{C}_{\text{Ising}}$ and $j \in \{0,1,\ldots,7\}$. In particular, the fusion coefficients are
\begin{equation}
N_{A_j B_k}^{C_l} = N_{AB}^{C} \delta_{l,[j+k]_8}
,
\end{equation}
where $N_{AB}^{C}$ are the Ising fusion coefficients of Eq.~(\ref{eq:Ising_N}) for $A,B,C \in \mathcal{C}_{\text{Ising}}$, and the fusion coefficients are only defined for $A_j , B_k , C_l \in \mathcal{C}$.

By solving the coherence conditions' polynomial equations, it has been shown~\cite{Bonderson07b,BondersonIP} that all consistent anyon models with this fusion algebra are obtained by taking the restricted product of anyon models $ \left. \text{Ising}^{(n)} \times \mathbb{Z}_8^{(w)} \right|_{\mathcal{C}}$. The $F$-symbols, $R$-symbols, quantum dimensions, and topological twists can be obtained by simply taking the product of their corresponding $\text{Ising}^{(n)}$ and $\mathbb{Z}_8^{(w)}$ values, e.g. $\theta_{A_j} = \theta_{A} \theta_j$. It was also found that there are $32$ distinct anyon models of this type and they are uniquely distinguished by the topological twist factors. Actually, this overcounts the number of distinct anyon models, because one is free to make the relabeling $I_2 = \psi_2^{\prime}$, $\psi_2 = I_2^{\prime}$, $I_6 = \psi_6^{\prime}$, and $\psi_6 = I_6^{\prime}$, which reduces the number of distinct anyon models to 16. With this count, we can see that the distinct anyon models can be uniquely distinguished by their value of $\theta_{\sigma_1}$ and can be described simply using $ \left. \text{Ising} \times \mathbb{Z}_8^{(w)} \right|_{\mathcal{C}}$, with $w = 0,\frac{1}{2},1,\ldots,\frac{15}{2}$. In fact, one can further reduce the count to 12 distinct anyon models, because, for $w\in \mathbb{Z}+\frac{1}{2}$, the anyon models with $w=w'(\text{mod }4)$ can be equated using the topological charge relabeling $\sigma_1 = \sigma_5^{\prime}$, $\sigma_3 = \sigma_7^{\prime}$, $\sigma_5 = \sigma_1^{\prime}$, and $\sigma_7 = \sigma_3^{\prime}$. We will, however, not equate models under this relabeling, because we wish to equate the $\mathbb{Z}_8$ (subscript) labels with electric charge (in units of $-e/4$ modulo $2e$). The minimal anyon models that describes the phase $X$ described in this paper are $X = \left. \text{Ising} \times \mathbb{Z}_8^{(w)} \right|_{\mathcal{C}}$ with $w=-\frac{1}{2}$ or $\frac{7}{2}$. We also note that the $\nu=1/2$ Moore-Read Pfaffian state is described by an anyon model of this form with $w=\frac{1}{2}$.

We can now write the $F$-symbols as
\begin{equation}
\label{eqn:combined-fusion}
\left[F^{A_j B_k C_l}_{D_{[j+k+l]_8}} \right]_{E_{[j+k]_8} F_{[k+l]_8} } = \left[ F^{A B C}_{D}\right]_{E F} e^{i \frac{\pi w}{4} j (k + l - [k+l]_{8} )}
,
\end{equation}
where $\left[ F^{A B C}_{D}\right]_{E F}$ are the corresponding Ising $F$-symbols.
Similarly, the $R$-symbols are
\begin{equation}
\label{eqn:combined-braiding}
R^{A_j B_k}_{C_{[j+k]_8}} = R^{A B}_{C} e^{i \frac{\pi w}{4} jk}
,
\end{equation}
where $R^{AB}_{C}$  are the corresponding Ising $R$-symbols.

It is worth reemphasizing that there is gauge freedom in specifying the $F$-symbols and $R$-symbols, so they can actually take different values while still representing the same physical model. The gauge invariant quantities, which uniquely distinguish these anyon models, are the fusion coefficients,
the quantum dimensions $d_{I_{2j}}=d_{\psi_{2j}}=1$ and $d_{\sigma_{2j+1}} = \sqrt{2}$, and the topological twist factors
\begin{eqnarray}
\theta_{I_{0}} &=& \theta_{I_{4}} = 1 ,\\
\theta_{\psi_{0}} &=& \theta_{\psi_{4}} = -1 ,
\end{eqnarray}
\begin{eqnarray}
\theta_{I_{2}} &=& \theta_{I_{6}} = e^{i \pi w} , \\
\theta_{\psi_{2}} &=& \theta_{\psi_{6}} = -e^{i \pi w} ,
\end{eqnarray}
\begin{eqnarray}
\theta_{\sigma_{1}} &=& \theta_{\sigma_{7}} = e^{i \frac{\pi}{8} (2w+1)} , \\
\theta_{\sigma_{3}} &=& \theta_{\sigma_{5}} = (-1)^{2w} e^{i \frac{\pi}{8} (2w+1)} .
\end{eqnarray}
The $S$-matrix is given by
\begin{equation}
S_{A_j B_k} = \frac{1}{2} S_{A B} e^{i \frac{\pi w}{2} j k }.
\end{equation}

When $w \in \mathbb{Z}$, we can see that $S = S_z \otimes S_{z'} \otimes S_{\text{Ising}}$, where $S_z = S_{z'}$ is the $S$-matrix of $\mathbb{Z}_2^{(0)}$, generated by $z = I_4$ and $z' = I_2$ when $w$ is even or $z'= \psi_2$ when $w$ is odd. In fact, since $z$ and $z'$ are bosons which are mutually local with all other topological charges, we can reduce the anyon model by identifying them with vacuum, i.e. using $\mathcal{Z} = \{z ,z'\}$. The anyon model that results from taking this quotient is Ising$^{(w)}$.

When $w \in \mathbb{Z} + \frac{1}{2}$, we can apply a $\mathbb{Z}_2$-grading using the fermion $z = \psi_4$, and the $S$-matrix takes the form $S = S_z \otimes \widehat{S}$, with
\begin{widetext}
\begin{equation}
\widehat{S} = \frac{1}{\sqrt{8}} \left[
\begin{array}{cccccc}
1  &  1  &  \sqrt{2} &  1  &  1  &  \sqrt{2} \\
1  &  1  &  -\sqrt{2} &  1  &  1  &  -\sqrt{2} \\
\sqrt{2}  &  -\sqrt{2}  & 0  &  e^{i \pi w} \sqrt{2}  &  -e^{i \pi w} \sqrt{2}  & 0 \\
1  &  1  &  e^{i \pi w} \sqrt{2} &  -1  &  -1  &  e^{i 3\pi w} \sqrt{2} \\
1  &  1  &  -e^{i \pi w} \sqrt{2} &  -1  &  -1  &  -e^{i 3\pi w} \sqrt{2} \\
 \sqrt{2}  &   - \sqrt{2}&  0 &  e^{i 3\pi w} \sqrt{2}   &  -e^{i 3\pi w} \sqrt{2}   &  0
\end{array}
\right]
\end{equation}
\end{widetext}
which is unitary. Hence, these anyon models correspond to $\mathbb{Z}_2$-graded TQFTs.

\subsection{$\mathbb{Z}_2$-graded modular extensions of $\left. \text{Ising} \times \mathbb{Z}_8^{(w)} \right|_{\mathcal{C}}$}
\label{sec:modular_extensions}

We consider $\mathbb{Z}_2$-graded modular extensions of the $\left. \text{Ising} \times \mathbb{Z}_8^{(w)} \right|_{\mathcal{C}}$ models with $w \in \mathbb{Z} + \frac{1}{2}$. These are modular anyon models that contain the original anyon model as a subsector and are $\mathbb{Z}_2$-graded in the sense that its set of topological charges can be written as $\tilde{\mathcal{C}} = \mathcal{C}_0 \oplus \mathcal{C}_1$, where the even sector's $\mathcal{C}_0 = \mathcal{C}$ is the charge set of the original anyon model, containing only anyons that are mutually local with the fermion $\psi_4$, and the odd sector's $\mathcal{C}_1$ is the set of extending charges, containing anyons whose braiding with the fermion $\psi_4$ results in a $-1$ phase. The total quantum dimension for the charges in the two sectors is equal $\mathcal{D}_0 = \mathcal{D}_1$.

One class of modular extensions is obtained by simply taking
\begin{equation}
\text{Ising}^{(n)} \times \mathbb{Z}_8^{(w-n)}
,
\end{equation}
where $n = 0,1,\ldots,7$, giving $8$ distinct extensions.
It is easy to see that the restriction to $\mathcal{C}$ of these extensions matches the original models (when $n$ is odd, this requires redefining $I_2 \leftrightarrow \psi_2$ and $I_6 \leftrightarrow \psi_6$), that the total quantum dimension squared $\mathcal{D}^2 = 32$ is twice that of the original models, and that the braiding of the extending anyons (i.e. those not in $\mathcal{C}$) with the fermion $\psi_4$ results in a $-1$ phase. The braiding and fusion properties are simply obtained by taking products, so we will not elaborate. These modular extensions have corresponding chiral central charges $c_{-} = \frac{1}{2} + n +(-1)^{w-n-\frac{1}{2}}$ (mod 8).

Another class of modular extensions is obtained by taking
\begin{equation}
\frac{\text{Ising}^{(m+n)} \times \text{Ising}^{(n)} \times \mathbb{Z}_8^{(w-n)}}{\left\langle (f,\psi,4) \right\rangle}
,
\end{equation}
where we denote the charge labels of the first Ising sector by $I$ (vacuum), $s$ (non-Abelian anyon), and $f$ (fermion). The denominator indicates moding out by the $\mathbb{Z}^{(0)}_2$ sector generated by the boson $(f,\psi,4)$, or, more physically, performing a condensation of this boson~\cite{Bais09}. In this case, the moding out amounts to identifying anyons by fusion with this boson, i.e. $(a,A,n) \sim (a\times f, A \times \psi , n+4)$, and removing the anyons that are not mutually local with this boson, i.e. those with $\theta_{(a,A,n)} \neq \theta_{(a\times f, A \times \psi , n+4)}$, from the resulting spectrum. This results in anyon models with 18 distinct topological charges
\begin{eqnarray}
\tilde{\mathcal{C}} &=& \left\{
(I,I,2j) , (I,\psi,2j) , (I,\sigma,2j+1) , \right.  \\
&& \left. (s,I,2j+1) ,(s,\sigma,0), (s,\sigma,2)
| j=0,1,2,3  \right\} \notag
.
\end{eqnarray}
The fusion and braiding properties are simply given by the product of those of $\text{Ising}^{(m+n)} \times \text{Ising}^{(n)} \times \mathbb{Z}_8^{(w-n)}$, up to the identification by fusion with the boson.
We will use the corresponding shorthand $I_{2j}$, $\psi_{2j}$, $\sigma_{2j+1}$, $s_{2j+1}$, $s\sigma_0$, and $s \sigma_2$. It is clear that the restriction to $\mathcal{C}$ results in the original anyon model and that the extending anyons aquire a $-1$ phase when braiding around the fermion $\psi_4$. These anyons have quantum dimensions $d_{I_{2j}} = d_{\psi_{2j}}=1$, $d_{\sigma_{2j+1}} = d_{s_{2j+1}} = \sqrt{2}$, and $d_{s\sigma_0} = d_{s\sigma_2}=2$. The total quantum dimension squared is $\mathcal{D}^2 = 32$. The twist factors of the new anyons are
\begin{eqnarray}
\theta_{s_{1}} &=& - \theta_{s_{3}} = -\theta_{s_{5}} = \theta_{s_{7}} = e^{i \frac{\pi}{8} \left( 2m + 2w + 1 \right) } \\
\theta_{s\sigma_0} &=&  e^{-i \pi (w-n)} \theta_{s\sigma_2} =  e^{i \frac{\pi}{4} \left( 2n + m + 1 \right) }
,
\end{eqnarray}
which gives 32 distinct models corresponding to the different values of $n=0,1,2,3$ and $m=0,1,\ldots,7$. (This count can be reduced to 8 through identifications involving topological charge redefinitions, but we do not consider this.) These modular extensions have corresponding chiral central charges $c_{-} = 1 +m + 2n +(-1)^{w-n-\frac{1}{2}}$ (mod 8).

We have not proven that these constitute all the $\mathbb{Z}_2$-graded modular extensions of the $\left. \text{Ising} \times \mathbb{Z}_8^{(w)} \right|_{\mathcal{C}}$ models with $w \in \mathbb{Z} + \frac{1}{2}$, but we believe they do. In any case, these appear to be the extensions obtained through combination with the 16 $\mathbb{Z}_2$-graded modular extensions of a free fermion~\cite{Kitaev06a}.


\end{document}